\documentclass[twocolumn,showpacs,preprintnumbers,amsmath,amssymb,%
prd,a4paper,floatfix,nofootinbib]{revtex4-1}  
\pdfoutput=1

\usepackage{graphics}
\usepackage{graphicx}
\usepackage{epsfig}
\usepackage{color}
\usepackage{hyperref}
\usepackage{latexsym}
\usepackage{amsmath}
\usepackage{amsthm}
\usepackage{amsfonts}
\usepackage{amssymb}
\usepackage{dsfont}
\usepackage{bm}
\usepackage{graphics,psfrag}
\usepackage{graphicx,psfrag}
\usepackage[tight]{units}

\newcommand{\eq}[1]{(\ref{#1})}
\newcommand{\fig}[1]{Fig.~{\ref{#1}}}
\newcommand{\tab}[1]{Tab.~{\ref{#1}}}
\newcommand{\FC}{\;,}
\newcommand{\FD}{\;.}
\newcommand{\RDH}[1]{{\mathrm{red}(#1)}}
\newcommand{\LMH}[1]{{\mathrm{lm}(#1)}}

\newcommand{\ov}{\overline}
\newcommand{\ot}{\frac{1}{2}}
\newcommand{\otf}{{\scriptstyle\frac{1}{2}}}
\newcommand{\ttf}{{\scriptstyle\frac{3}{2}}}

\begin{document}
\date{\today}

\title{Symmetries of hadrons after unbreaking the chiral symmetry}
\author{L.~Ya.~Glozman}
\email{leonid.glozman@uni-graz.at}
\affiliation{Institut f\"ur Physik, FB Theoretische Physik, Universit\"at
Graz, A--8010 Graz, Austria}
\author{C.~B.~Lang}
\email{christian.lang@uni-graz.at}
\affiliation{Institut f\"ur Physik, FB Theoretische Physik, Universit\"at
Graz, A--8010 Graz, Austria}
\author{M.~Schr\"ock}
\email{mario.schroeck@uni-graz.at}
\affiliation{Institut f\"ur Physik, FB Theoretische Physik, Universit\"at
Graz, A--8010 Graz, Austria}

\begin{abstract}
We study hadron correlators upon artificial restoration of the
spontaneously broken chiral symmetry.  In a dynamical lattice simulation
we remove  the lowest lying eigenmodes of the Dirac operator from the
valence quark propagators and study evolution of the  hadron masses
obtained. All mesons and baryons in our  study, except for a pion,
survive unbreaking the chiral symmetry and their exponential decay
signals become essentially better. From the analysis of the observed
spectroscopic patterns we conclude that confinement still persists while
the chiral symmetry is restored. All hadrons fall into different chiral
multiplets. The broken $U(1)_A$ symmetry does not get restored upon
unbreaking the chiral symmetry. We also  observe signals of some higher
symmetry that includes chiral symmetry as a subgroup.  Finally, from
comparison of the $\Delta - N$ splitting before and after unbreaking of
the chiral symmetry we conclude that both the color-magnetic and the
flavor-spin quark-quark interactions are of equal importance.
\end{abstract}
\maketitle

\section{Introduction}

Highly excited hadrons in the $u,\,d$ sector reveal some parity doubling
\cite{Glozman:1999tk,Cohen:2001gb,Cohen:2002st,
Glozman:2002cp,Glozman:2003bt,Glozman:2007ek,Glozman:2007ek,
Jaffe:2006aq,Jaffe:2005sq,Jaffe:2006jy, Shifman:2007xn} and possibly some
higher symmetry. It was  conjectured that this parity  doubling reflects
effective restoration of chiral symmetry, i.e., insensitivity of the
hadron mass generation mechanism to the effects of chiral symmetry
breaking in the vacuum \cite{Glozman:1999tk,Cohen:2001gb,Cohen:2002st,
Glozman:2002cp,Glozman:2003bt,Glozman:2007ek}. Whether this conjecture is
correct or not can be answered experimentally since the conjectured
symmetry requires existence of some  not yet observed states.

Recent and most complete experimental analysis  on highly excited
nucleons that includes not only elastic $\pi N$, but also the
photoproduction data, does report  evidence for some of the missing
states and the parity doubling patterns look  now even better than before
\cite{Anisovich:2011sv}. 

The question of a possible symmetry in hadron spectra is one of the
central questions for QCD since it would help to understand dynamics of
confinement and chiral symmetry breaking as well as their role for the
hadron mass generation.

Another ``experimental'' tool to address the issue of the hadron mass
generation is lattice QCD. Equipped with the QCD Lagrangian and
Monte-Carlo techniques, one can calculate, at least in principle, hadron
masses and  other hadron properties from first principles. Enormous
progress has been achieved for the hadron ground states. The problem of
excited states, especially above the multihadron thresholds like $\pi N$,
$\Delta \pi$, $\pi \pi$,  $\pi \rho$, \ldots turns out to be much more
difficult and demanding than was initially anticipated. When it is solved
lattice results should reproduce experimental patterns and possibly
indicate some still missing states.

Still, the  mass of a hadron by itself, obtained from the experiment or
from the lattice simulations, tells us not so much about the physics
which is behind the mass generation. The pattern of all hadrons, on the
contrary, could shed some light on the underlying  dynamics if there are
some obvious symmetries in the pattern or if its  regularities can be
systematically explained.

The most interesting issue is to get some insight on how QCD ``works'' in
some important cases and understand the underlying physical picture. In
this sense one can use lattice QCD as a tool to  explore the
interrelations between confinement and chiral symmetry breaking. In
particular, we can ask the question whether hadrons and confinement will
survive after having artificially removed the quark condensate of the
vacuum.  This can be achieved via removal of the low-lying eigenmodes of
the Dirac operator, which is a well defined procedure
\cite{DeGrand:2003sf,Cohen:2006bq}. 

In the past mainly the opposite was explored. After suggestions within the 
instanton liquid model \cite{Schafer:1996wv} the effect of the low-lying
chiral modes on the $\rho$ and other correlators was studied on the lattice.
In a series of papers \cite{DeGrand:2000gq,Neff:2001zr,
DeGrand:2001tm,DeGrand:2003sf} it
was  shown that low modes saturate the pseudoscalar and axial vector
correlators at large distances and do not affect the part where high-lying
states appear.  In \cite{DeGrand:2003sf,DeGrand:2004qw} low mode saturation 
and also effects of low mode removal for mesons were studied for
quenched configurations with the overlap Dirac operator 
\cite{Neuberger:1997fp,Neuberger:1998wv}.
Subsequently low modes were utilized to improve the convergence of the
determination  of hadron propagators
\cite{DeGrand:2003sf,DeGrand:2004qw,DeGrand:2004wh,Giusti:2004yp,
Bali:2009hu,Bali:2010se} studying the efficiency when using
the low modes of the Dirac operator or the Hermitian Dirac operator. 

We are studying the complementary case, i.e., removal of the low modes and
we will refer to this as ``unbreaking''
the chiral symmetry. This issue has been addressed in a recent paper
\cite{Lang:2011qy,Lang:2011ai} where the low-lying eigenmodes of the
Dirac operator have been removed from the quark Green's function and 
masses of the lowest mesons  $\pi,\rho,a_0$ and $a_1$ have been
calculated with such truncated quark propagators.
The truncated Landau gauge quark propagator itself has been investigated
in \cite{Schrock:2011hq} where the loss of dynamical mass generation
in the infrared sector of the propagator has been demonstrated.

After the unbreaking of the chiral symmetry the signal from the
$\pi$-meson, obtained with the pseudoscalar quark-antiquark operator,
disappears, which is consistent with the (pseudo) Goldstone boson nature
of the pion. Indeed, with the artificially restored chiral symmetry there
cannot be Goldstone bosons. What is very interesting, is that other
low-lying mesons survive and the quality of their signals become even
essentially better after extraction of the low-lying eigenmodes of the
Dirac operator,  responsible for chiral symmetry breaking. The very fact
that hadrons survive the unbreaking of the chiral symmetry tells that
there is confinement in the system even without  the quark condensate. (A
similar behavior was found in \cite{Suganuma:2011kn,Gongyo:2012vx}, where
the effect of such a removal on the static quark potential was studied.)

After extraction of the quark condensate the lowest-lying $\rho$ and
$a_1$  mesons demonstrate restoration of the chiral symmetry - they 
become degenerate - and their mass is rather large. This disproves a
rather popular  assertion that, e.g., the $\rho$-mass is entirely due to
the  quark condensate of the vacuum. As a physical implication  we should
then not expect a drop off the $\rho$-mass in a dense medium, which is a
very popular issue both theoretically and experimentally for the last two
decades \cite{Brown:1991kk}. This result should also be of importance for
a debated issue of confining but chirally symmetric matter at low
temperatures and large density \cite{Glozman:2007tv,Glozman:2008fk,
Glozman:2009sa,Glozman:2011eu}.

The conclusion of \cite{Lang:2011qy} about  survival of confinement after
unbreaking of the chiral symmetry was obtained on the limited basis of
the lowest-lying mesons. In order to see it more clearly we need to
extend the number of extracted states, in particular to include radially
and orbitally excited hadrons (following quark model terminology).
Consequently, we now add the $b_1$ and $\rho'$ states to the above
mentioned list of mesons. We limit ourselves only to the isovector mesons
since the isoscalar mesons would require inclusion of disconnected
graphs, which is numerically very costly.

Most importantly, we study behavior of the ground and excited positive
and negative parity states in the $N$ and $\Delta$ spectra. The baryonic
states add additional information about  existence or nonexistence of
confinement. They do allow to see that confinement does survive after the
restoration of chiral symmetry. Second, we  see some traces of the higher
symmetry, higher than simply $SU(2)_R \times SU(2)_L$. This observation
may be related with the higher symmetry seen in the highly excited
hadrons.

Our paper is organized as follows. In the next section we remind on the
connection between low eigenmodes of the Dirac operator and the quark
condensate in the vacuum. We discuss some basic aspects of removal of
Dirac eigenmodes from the quark propagator. In Sect. \ref{sec:setup} we 
present the details of the lattice simulation. The fourth section is
devoted to the description of the baryon and meson interpolators used for
our study of hadrons. In Sect. \ref{sec:results} we show and analyze the
results and draw conclusions. Finally, in the last section, we  briefly
summarize our main observations.

\section{The quark condensate and the Dirac operator}\label{sec:basics}

The lowest eigenmodes of the Dirac operator  are related (in the chiral
limit) to the quark condensate of the vacuum.  This is encoded in the
Banks--Casher  relation \cite{Banks:1979yr}
\begin{equation}\label{BC}
< 0 | \bar q q | 0 > = - \pi \rho(0)\FC
\end{equation}
\noindent
where $\rho(0)$ is a density of the lowest quasi-zero eigenmodes of the
Dirac operator. Here the sequence of limits is important: first, the
infinite volume limit  at finite quark  mass is assumed and then the
chiral limit should be taken. The opposite sequence would produce no
chiral condensate as in the finite volume there cannot be any spontaneous
breaking of chiral symmetry.

All lattice calculations are performed on a lattice of a finite volume.
In the finite lattice volume the spectrum of the Dirac operator is
discrete and the energy of the lowest nonzero mode of the Dirac operator
is finite. Consequently, the quark condensate is strictly speaking zero.
However, increasing the lattice volume the gap in the spectrum of the
eigenmodes becomes smaller and smaller and the density of the lowest nonzero
eigenmodes increases. A well defined limit of this density scaling
exists: the number of such eigenvalues
in a given interval adjacent to the real axis scales with the lattice volume. 
(In \cite{Osborn:1998nm} is is argued\footnote{We thank Kim Splittorff for pointing us
to that reference} that the number of relevant eigenvalues
should scale proportional the square root of the number of lattice points.)

In \cite{Bali:2010se,Lang:2011qy} it was established that the eigenmodes
of the Hermitian Dirac operator $D_5 \equiv \gamma_5\,D$ result in a
faster saturation of the pseudoscalar correlator when approximating quark
propagators by the lowest eigenmodes only, compared to the eigenmodes of
the Dirac operator $D$. Therefore, we focus on reducing the quark
propagators in terms of eigenmodes of $D_5$ rather than $D$.

From the lattice calculations in a given finite volume we cannot
say a priori which and how many lowest eigenmodes of the Dirac operator are
responsible for the quark condensate of the vacuum. For the overlap operator the
real eigenvalues correspond to exact chiral modes, the zero modes (instantons);
for Wilson-type operators one may associate real eigenvalues with zero modes.
Their weight is suppressed in the infinite volume limit. The
Banks-Casher relation, however, relies only on the density of nonzero modes. 
For the Hermitian Dirac operator $D_5$ there is no simple method to distinguish
between the real modes of the $D$ and its small, but complex modes.
In \cite{Lang:2011qy,Lang:2011ai} we discuss the integral over the
distribution of the (real) eigenvalues of $D_5$. There we observe a transition
region up to roughly twice the size of the quark mass  corresponding to
$\mathcal{O}(16-32)$  eigenmodes, as also observed in, e.g., 
\cite{Luscher:2007se,Giusti:2008vb,Necco:2011vx,Splittorff:2011bj}. 

We follow the
procedure  to remove an increasing number of the lowest Dirac modes and study the
effects of the (remaining) chiral symmetry breaking on the  masses of hadrons.
To be specific, we construct reduced quark propagators
\begin{equation}\label{eq:red5}
S_\RDH{k}=S-S_\LMH{k}\equiv S- 
\sum_{i\le k} \mu_i^{-1} |{v_i}\rangle\langle{v_i}|\gamma_5\FC
\end{equation}
where $S$ is the standard quark propagator obtained from the inversion of
the Dirac operator, the $\mu_i$ are the (real) eigenvalues of $D_5$,
$|{v_i}\rangle$ are the corresponding eigenvectors and $k$ represents the
reduction parameter which will be varied from $0-128$.

Note that the low-mode contribution (lm) of the quark propagators,
$S_\LMH{k}$, must act on the same quark sources as $S$, see discussion
below.

\section{The setup}\label{sec:setup}
\subsection{Dirac operator}

For the dynamical quarks of our configurations as well as for the valence
quarks of our study the so called chirally improved (CI) Dirac operator
\cite{Gattringer:2000js,Gattringer:2000qu} has been used. The latter
represents an approximate solution to the Ginsparg--Wilson equation and
therefore offers better chiral properties than the Wilson Dirac operator
while being less expensive, in terms of computation time, in comparison
to the  chirally exact overlap operator.

\subsection{Gauge configurations}

We performed our study on 161 gauge field configurations 
\cite{Gattringer:2008vj,Engel:2010my} that were generated for two
degenerate dynamical light CI fermions with a corresponding  pion mass
$m_\pi=\unit[322(5)]{MeV}$. The lattice size is $16^3\times 32$ and the
lattice spacing $a=\unit[0.144(1)]{fm}$. 

\subsection{Quark source smearing}\label{sec:smear}

In order to obtain quark propagators, the Dirac operator has to be
inverted on given quark sources. To improve the signal in hadron
correlators, extended sources of Gaussian form
\cite{Gusken:1989ad,Best:1997qp} instead of point sources are being used.
Using several different extended
sources allows for a larger operator basis in the variational method
\cite{Luscher:1990ck,Michael:1985ne,Burch:2004he}. We use three different kinds of sources:  narrow
(\unit[0.27]{fm}) and  wide (\unit[0.55]{fm})  sources, which are
approximately of Gaussian shape, and a derivative source.

The narrow (wide) sources will be denoted by a subscript $n$ ($w$) of the
quark fields  and the derivative source by $\partial_i$, respectively. 
The details of the calculation of the smeared quark sources are given in 
\cite{Engel:2010my}.

\subsection{Variational method}\label{sec:varmeth}

In order to disentangle the excited states from the ground state (and also 
to provide cleaner signals for the ground states) we use the
variational method \cite{Luscher:1990ck,Michael:1985ne}. One computes 
cross-correlators $C_{ik}(t)=\langle  O_i(t) O_k(0)^\dagger\rangle$ between 
several different lattice interpolators and solves the
generalized eigenvalue problem
\begin{equation}
C(t) \vec u_n(t)=\lambda_n(t) C(t_0) \vec u_n(t)\;,
\end{equation}
in order to approximately recover the energy eigenstates $|n\rangle$.
The eigenvalues allow us to get the energy values 
$\lambda_n(t)\sim\exp(-E_n\,t)$ and the eigenvectors serve as 
fingerprints of the states, indicating their content in terms of the 
lattice interpolators. In our plots we show $\lambda_n(t)$, the 
effective masses $E_n(t)=\log(\lambda_n(t)/\lambda_n(t+1))$ and the $t$-
dependence of the eigenvectors in order to verify the state identification.
The quality of the results depends on the statistics and
the provided set of lattice operators 
(for a discussion see \cite{Engel:2010my}). The used interpolators are 
discussed in Sect.\ \ref{sec:interpolators}.

\subsection{Dirac eigenmodes}

On the given gauge field configurations we calculated the lowest 128
eigenmodes of the Hermitian Dirac operator $D_5$ using ARPACK which is an
implementation of the Arnoldi method to calculate a part of the spectrum of
arbitrary matrices \cite{Lehoucq:1998xx}.

Once the eigenmodes have been calculated and the quark propagators $S$
have been obtained  by inverting the Dirac operator on the three types of
sources mentioned in Sec.~\ref{sec:smear},  we can construct the reduced
propagators $S_\RDH{k}$ according to \eq{eq:red5}  after multiplying the
low-mode part of the propagator,  $S_\LMH{k}$, with the same three source
types, respectively.

\section{Hadron interpolators}\label{sec:interpolators}

Here we list the baryons and mesons we studied under Dirac low-mode
reduction and give the interpolating fields for each individual.

\subsection{Baryons}
We analyze the nucleon and $\Delta$ baryons both with positive and 
negative parity. For the interpolators we use Gaussian smeared quark
sources ($n$ and $w$).  For the nucleon we adopt three different Dirac
structures, resulting in 18 interpolators (see
\tab{tab:baryoninterpolators1}) where we left out those operators that
are similar to other ones due to  isospin symmetry. The construction of
the nucleon interpolators is given by
\begin{equation}\label{eq:nucleon_def}
N^{(i)} = \epsilon_{abc}\, \Gamma_1^{(i)}\, u_a\, \big( u_b^T\, \Gamma_2^{(i)}\, d_c -
d_b^T\, \Gamma_2^{(i)}\, u_c \big)\FD
\end{equation}
For the $\Delta$,
\begin{equation}\label{eq:delta_def}
\Delta_k = \epsilon_{abc}\, u_a\, \big(u_b^T\, C\, \gamma_k\, u_c \big)\FC
\end{equation}
we use only one Dirac structure and the six corresponding interpolators
are listed in \tab{tab:baryoninterpolators2}. We use parity projection
for all baryons and Rarita-Schwinger projection for the $\Delta$
\cite{Engel:2010my}. The sink interpolators are also projected to zero
spatial momentum.

\begin{table}[tb]
\begin{center}
\begin{tabular}{ccc|c|c}
\hline
\hline
$\chi^{(i)}$ & $\Gamma^{(i)}_1$ & $\Gamma^{(i)}_2$ & smearing &
\#$_N$ \\
\hline
{$\chi^{(1)}$} &
{$\mathds{1}$} &
{$C\,\gamma_5$} & $(nn)n$ & 1 \\
 & & & $(nn)w$ & 2  \\
 & & & $(nw)n$ & 3  \\
 & & & $(nw)w$ & 4  \\
 & & & $(ww)n$ & 5  \\
 & & & $(ww)w$ & 6  \\
\hline
{$\chi^{(2)}$} &
{$\gamma_5$} &
{$C$} & $(nn)n$ & 7  \\
 & & & $(nn)w$ & 8   \\
 & & & $(nw)n$ & 9   \\
 & & & $(nw)w$ & 10  \\
 & & & $(ww)n$ & 11  \\
 & & & $(ww)w$ & 12  \\
\hline
{$\chi^{(3)}$} &
{$ i\,\mathds{1}$} &
{$C\,\gamma_t\,\gamma_5$} & $(nn)n$ & 13  \\
 & & & $(nn)w$ & 14  \\
 & & & $(nw)n$ & 15  \\
 & & & $(nw)w$ & 16  \\
 & & & $(ww)n$ & 17  \\
 & & & $(ww)w$ & 18  \\
\hline
\hline
\end{tabular}
\end{center}
\caption{Interpolators for the $N$ channel. The Dirac structures
$\chi^{(i)}$, the quark smearings and the corresponding interpolator
numbers \#$_N$ are given.}
\label{tab:baryoninterpolators1}
\end{table}

\begin{table}[tb]
\begin{center}
\begin{tabular}{c|c}
\hline
\hline
smearing & \#$_\Delta$ \\
\hline
$(nn)n$ & 1 \\
$(nn)w$ & 2 \\
$(nw)n$ & 3 \\
$(nw)w$ & 4 \\
$(ww)n$ & 5 \\
$(ww)w$ & 6 \\
\hline
\hline
\end{tabular}
\end{center}
\caption{Interpolators for the $\Delta$ channel. The quark smearings and
the corresponding interpolator numbers \#$_\Delta$ are given.}
\label{tab:baryoninterpolators2}
\end{table}

\subsection{Mesons}
We investigate isovector mesons of spin 1.  Isoscalars require the
calculation of disconnected graphs which are computationally too
demanding for the type of fermion action used.  The scalar meson $a_0$ 
as well as the pseudoscalar  pion  was studied already in
\cite{Lang:2011qy}.

Thus, the studied nonexotic channels are the $J^{PC}$ combinations
$1^{--}$ ($\rho$), $1^{++}$ ($a_1$) and $1^{+-}$ ($b_1$). For the
ana\-ly\-sis of the mesons we include derivative sources
\cite{Gattringer:2008be} in the construction of the interpolators to
provide a large operator basis for the variational method. In Table
\ref{tab:interpols} we list only those  interpolators explicitly whose
combination resulted in a good signal in practice when plugged into the
variational method. The sink interpolators are projected to zero spatial
momentum. A more complete list of possible interpolating fields is given
in  \cite{Engel:2010my,Engel:2011aa}. 

\begin{table}[tb]
\begin{center}
\begin{tabular}{c|c}
\hline
\hline
\#$_\rho$ & interpolator(s) \\
\hline
 1 & $\ov{a}_n \gamma_k b_n$ \\
 8 & $\ov{a}_w \gamma_k\gamma_t b_w$ \\
12 & $\ov{a}_{\partial_k} b_w - \ov{a}_w b_{\partial_k}$ \\
17 & $\ov{a}_{\partial_i} \gamma_k b_{\partial_i}$ \\
22 & $\ov{a}_{\partial_k} \epsilon_{ijk} \gamma_j\gamma_5 b_w - \ov{a}_w \epsilon_{ijk} \gamma_j\gamma_5 b_{\partial_k}$ \\
\hline
\#$_{a_1}$ & interpolator(s) \\
\hline
 1 & $\ov{a}_n \gamma_k \gamma_5 b_n$ \\
 2 & $\ov{a}_n \gamma_k \gamma_5 b_w+\ov{a}_w \gamma_k \gamma_5 b_n$ \\
 4 & $\ov{a}_w \gamma_k \gamma_5 b_w$ \\
\hline
\#$_{b_1}$ & interpolator(s) \\
\hline
 6 & $\ov{a}_{\partial_k} \gamma_5 b_n-\ov{a}_n \gamma_5 b_{\partial_k}$ \\
 8 & $\ov{a}_{\partial_k} \gamma_5 b_w-\ov{a}_w \gamma_5 b_{\partial_k}$ \\
\hline
\hline
\end{tabular}
\caption{Interpolators for (top) the $\rho$-meson, $J^{PC}=1^{--}$,
(middle)  the $a_1$-meson, $J^{PC}=1^{++}$, and (bottom) the $b_1$-meson,
$J^{PC}=1^{+-}$. The first column shows the number, the second shows the
explicit form of the interpolator. The numbers refer to the
classification in \cite{Engel:2011aa}. 
}
\label{tab:interpols}
\end{center}
\end{table}

\begin{figure*}[tb]
  \hspace*{24pt}
  \includegraphics[width=0.4\textwidth]{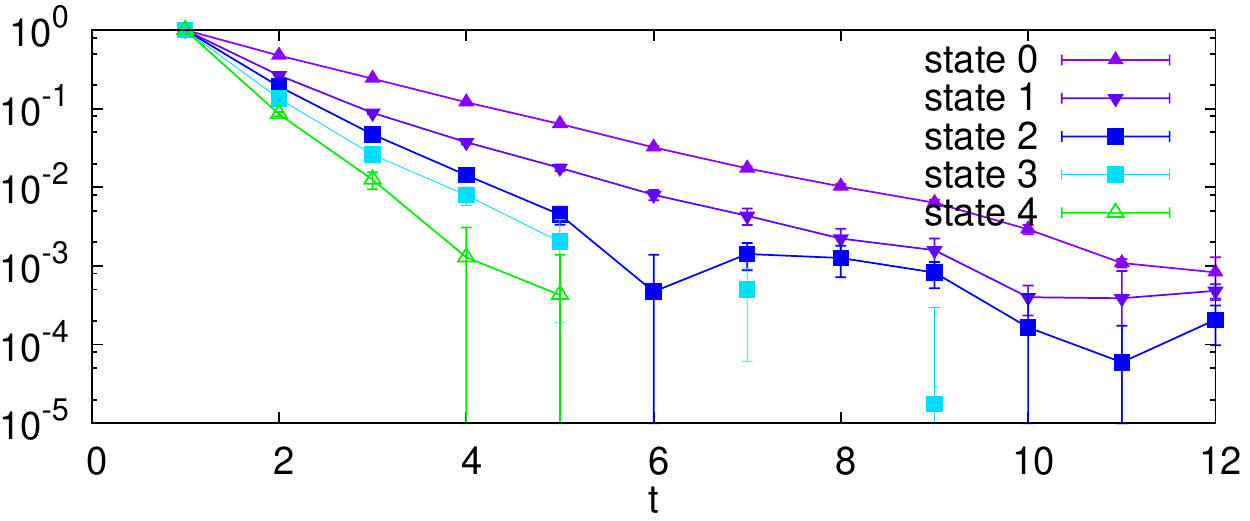}\hfill
  \includegraphics[width=0.4\textwidth]{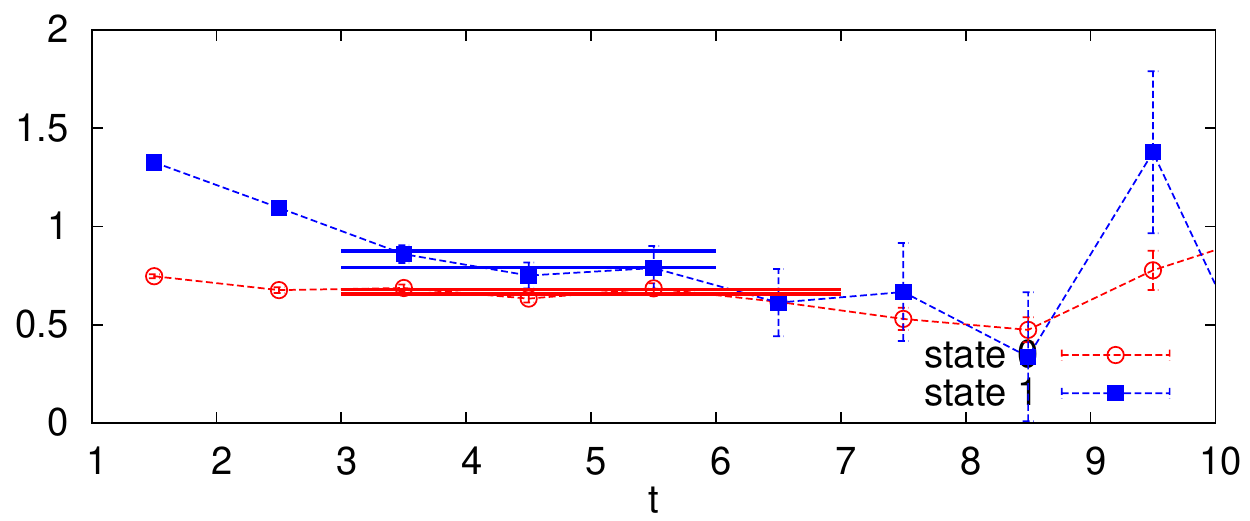}
  \hspace*{24pt}\hfil\\
  \hspace*{24pt}
  \includegraphics[width=0.4\textwidth]{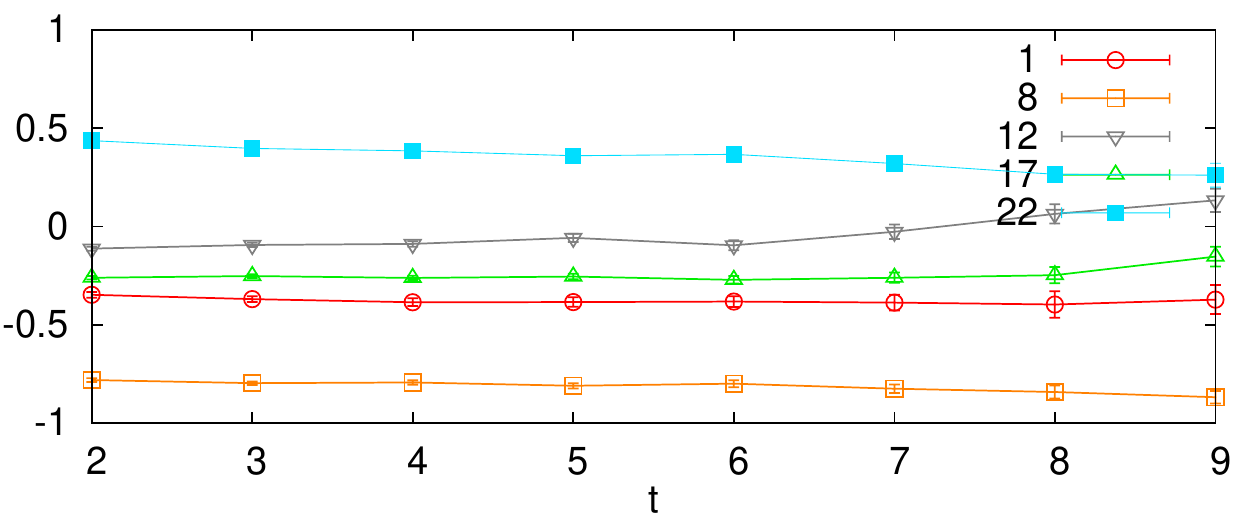}\hfill
  \includegraphics[width=0.4\textwidth]{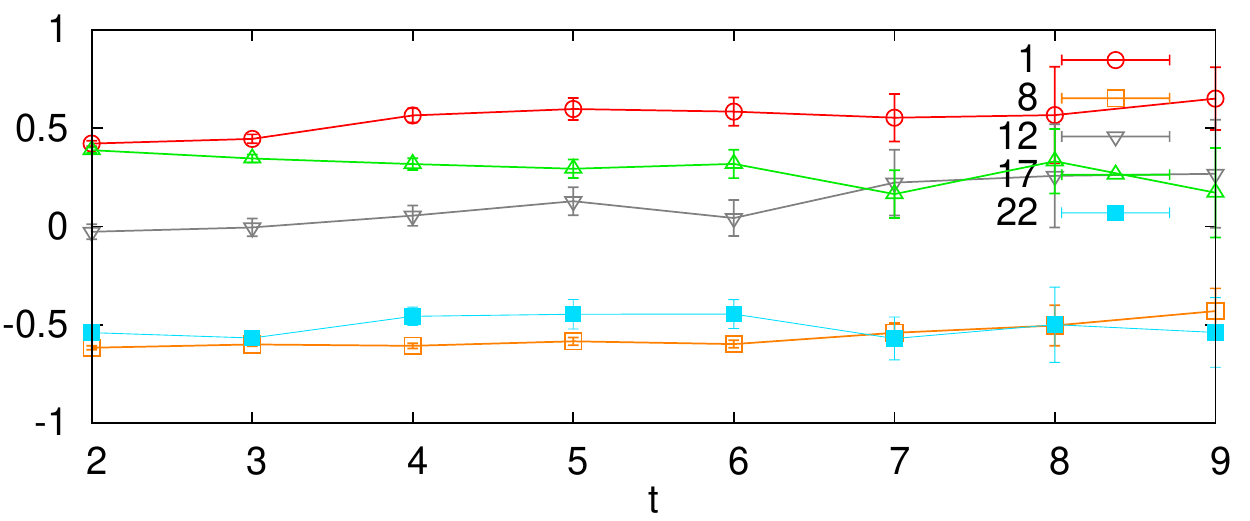}
  \hspace*{24pt}\hfil\\
  \caption{$\rho$ with 12 eigenmodes subtracted: The correlators for all eigenstates (upper left),
   effective mass plot for the two lowest states (upper right),    
   eigenvectors corresponding to the ground state (lower left).
   and 1st excited state (lower right). 
}\label{fig:rho_1}
\end{figure*}

\begin{figure*}[tb]
  \hspace*{24pt}
  \includegraphics[width=0.4\textwidth]{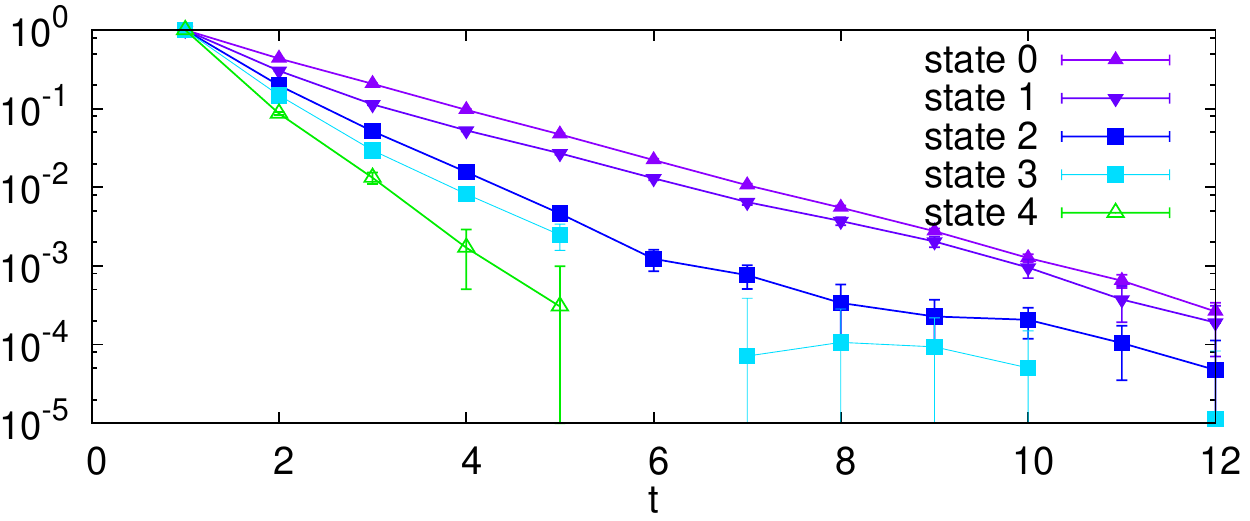}\hfill
  \includegraphics[width=0.4\textwidth]{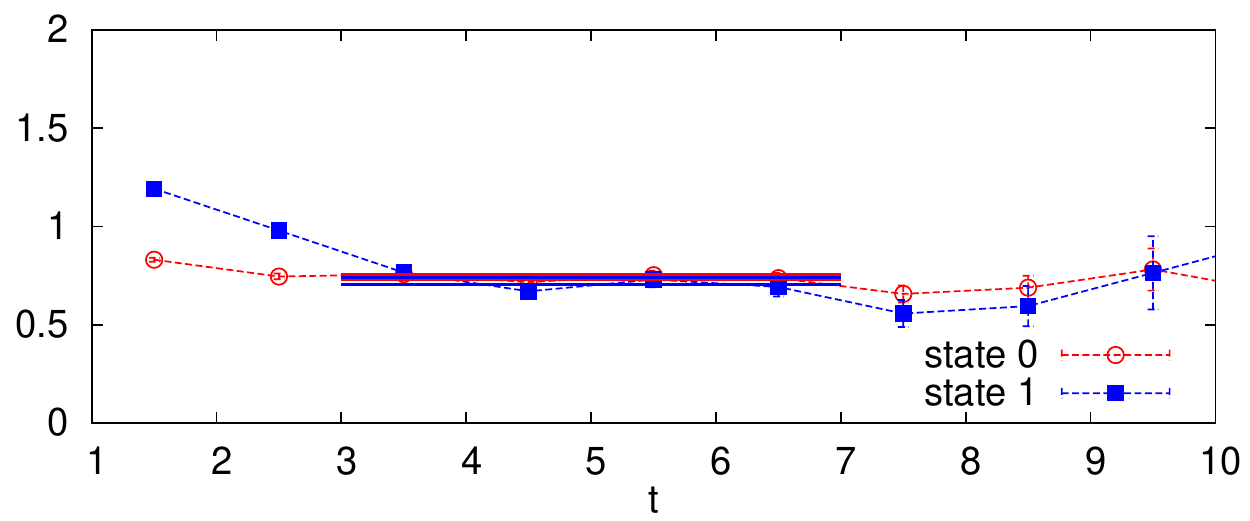}
  \hspace*{24pt}\hfil\\
  \hspace*{24pt}
  \includegraphics[width=0.4\textwidth]{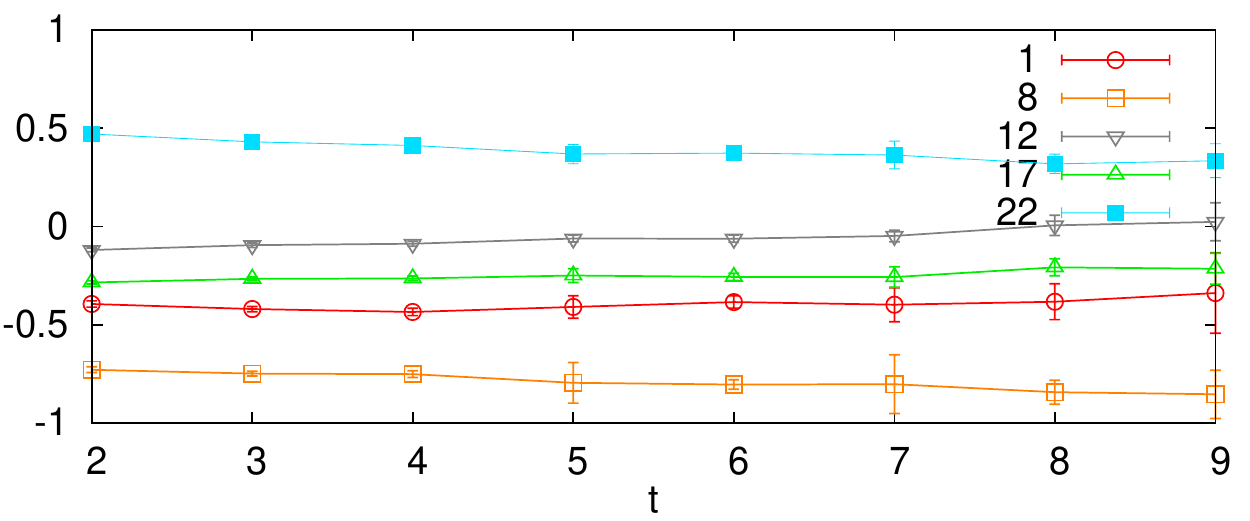}\hfill
  \includegraphics[width=0.4\textwidth]{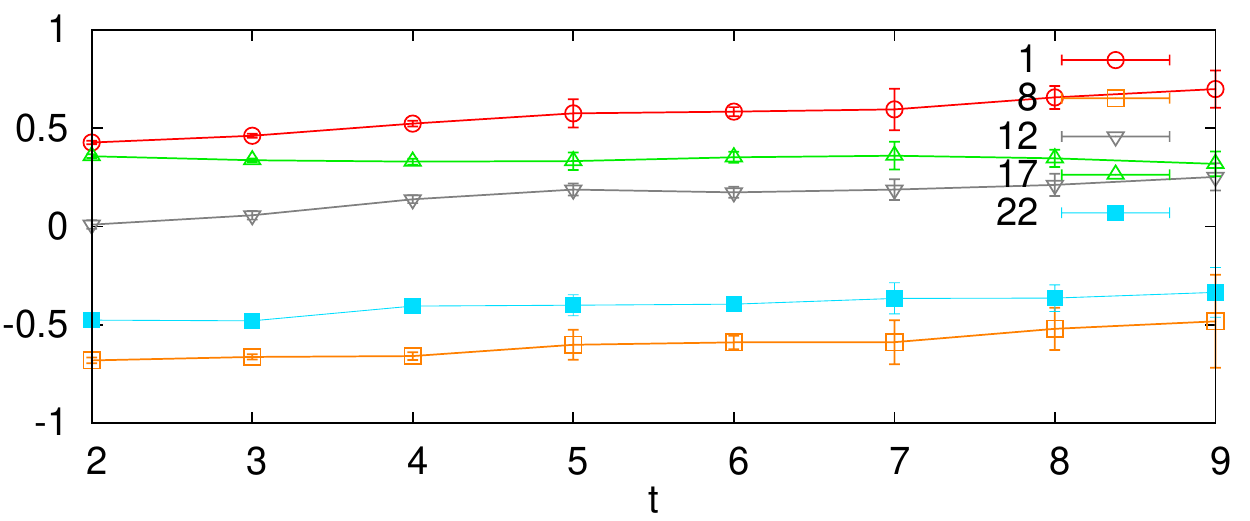}
  \hspace*{24pt}\hfil\\
  \caption{$\rho$ with 32 eigenmodes subtracted: The correlators for all eigenstates (upper left),
   effective mass plot for the two lowest states (upper right),    
   eigenvectors corresponding to the ground state (lower left).
   and 1st excited state (lower right). 
}\label{fig:rho_2}
\end{figure*}

\begin{figure*}[tb]
  \hspace*{24pt}
  \includegraphics[width=0.4\textwidth]{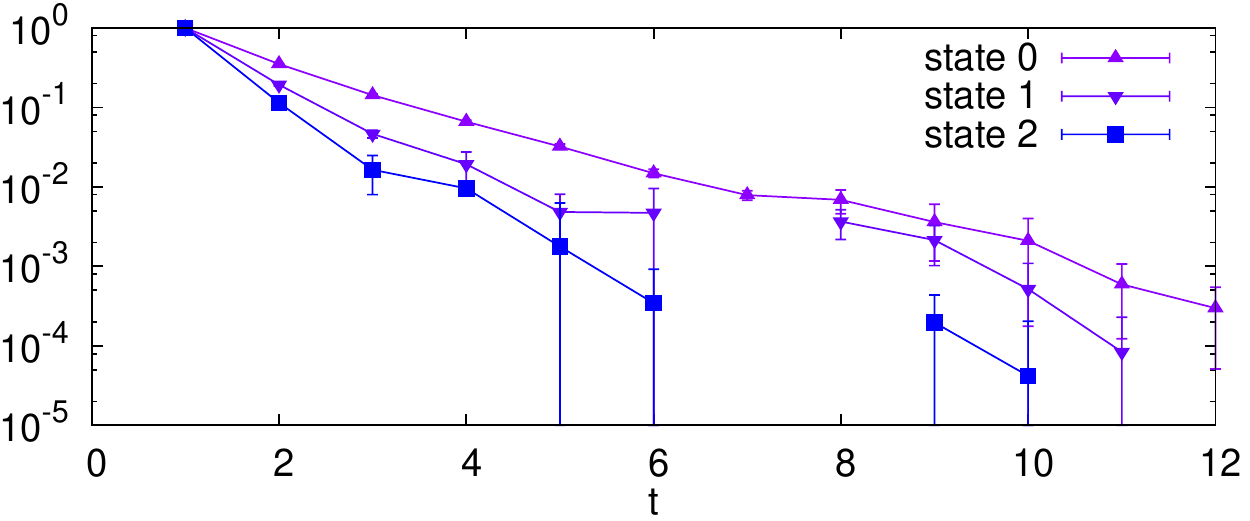}\hfill
  \includegraphics[width=0.4\textwidth]{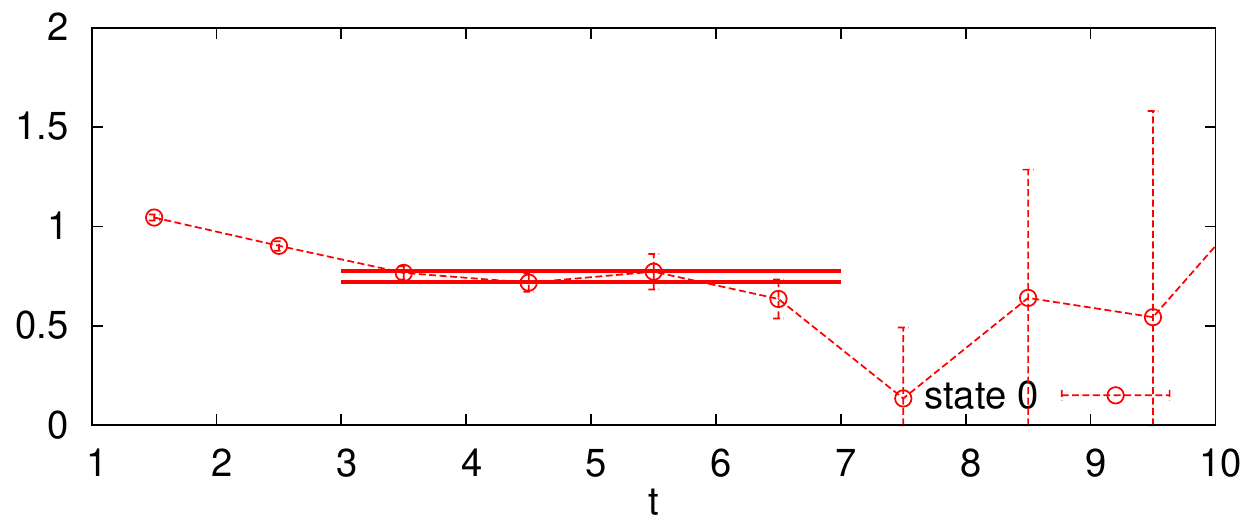}
  \hspace*{24pt}\hfil\\
  \hspace*{24pt}
  \includegraphics[width=0.4\textwidth]{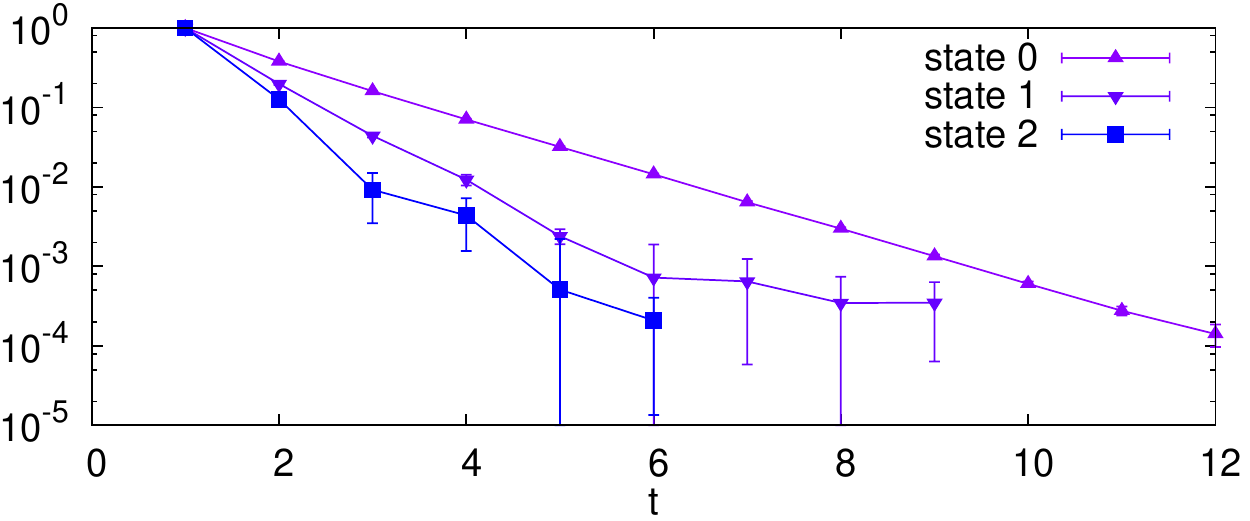}\hfill
  \includegraphics[width=0.4\textwidth]{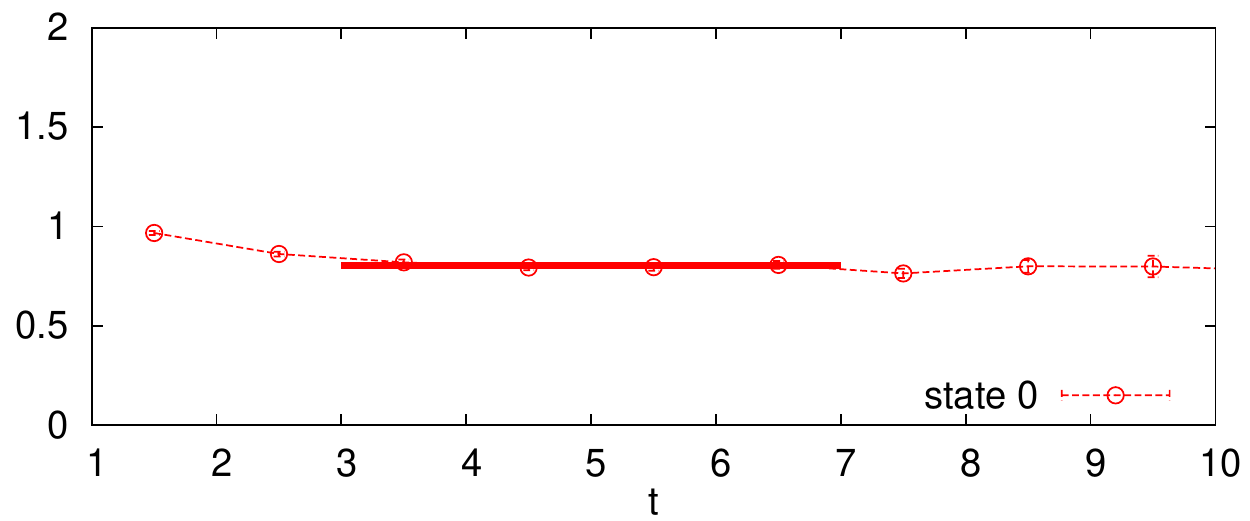}
  \hspace*{24pt}\hfil\\
  \caption{$a_1$ with 4 (upper row) and 64 (lower row) eigenmodes subtracted:
   The correlators for all eigenstates (left),
   effective mass plot for the lowest state (right).
}\label{fig:a1_1}
\end{figure*}

\begin{figure*}[tb]
  \hspace*{24pt}
  \includegraphics[width=0.4\textwidth]{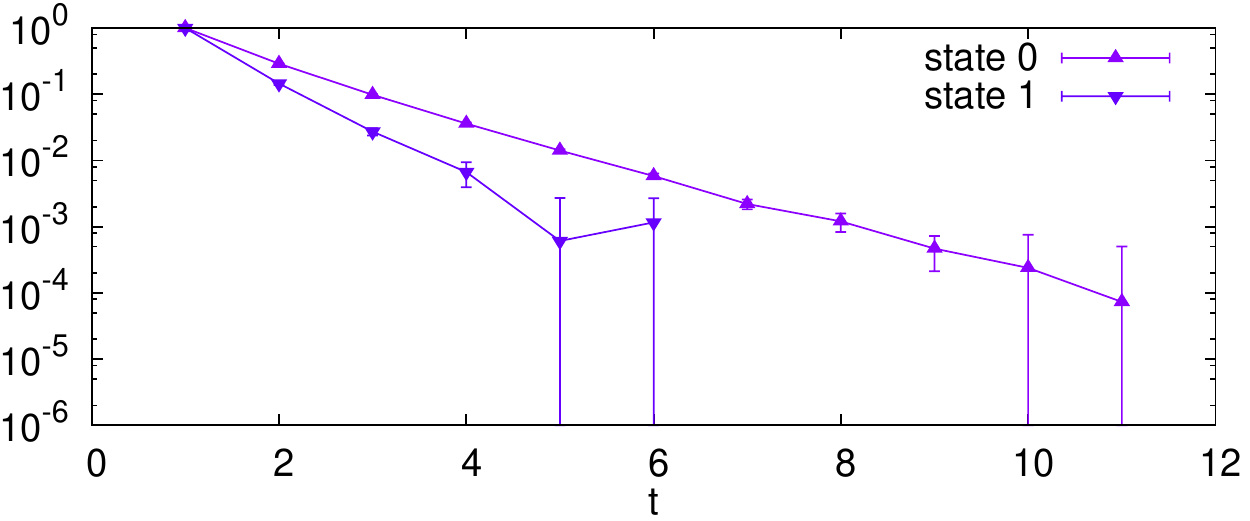}\hfill
  \includegraphics[width=0.4\textwidth]{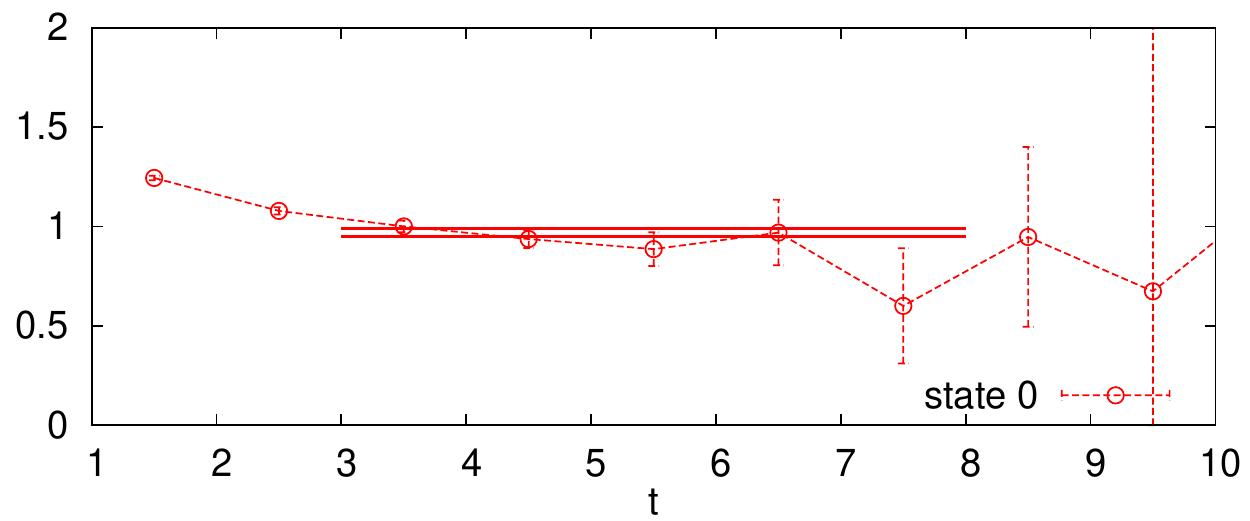}
  \hspace*{24pt}\hfil\\
  \hspace*{24pt}
  \includegraphics[width=0.4\textwidth]{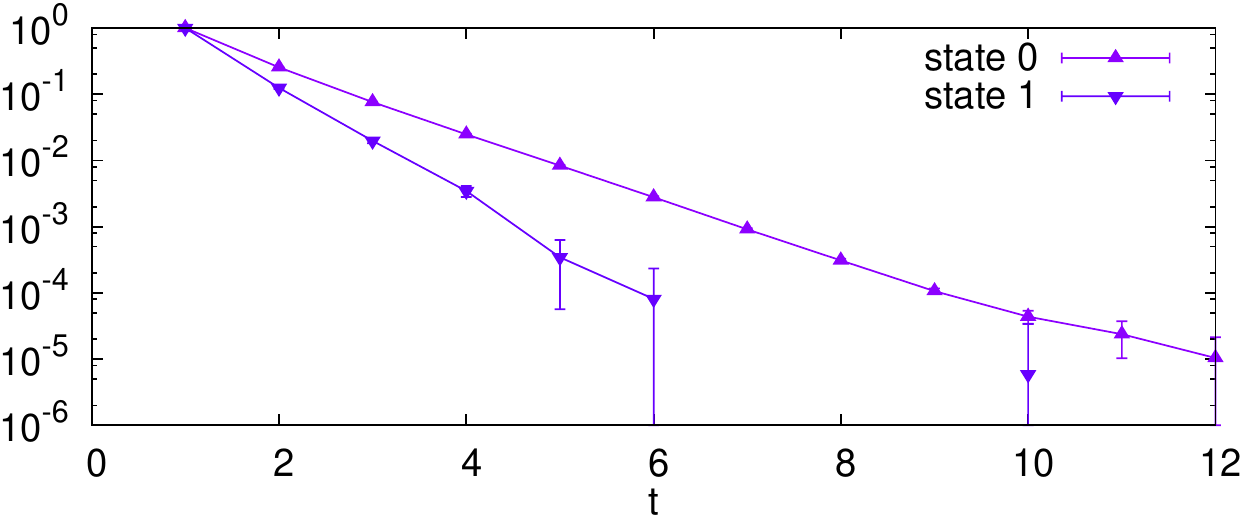}\hfill
  \includegraphics[width=0.4\textwidth]{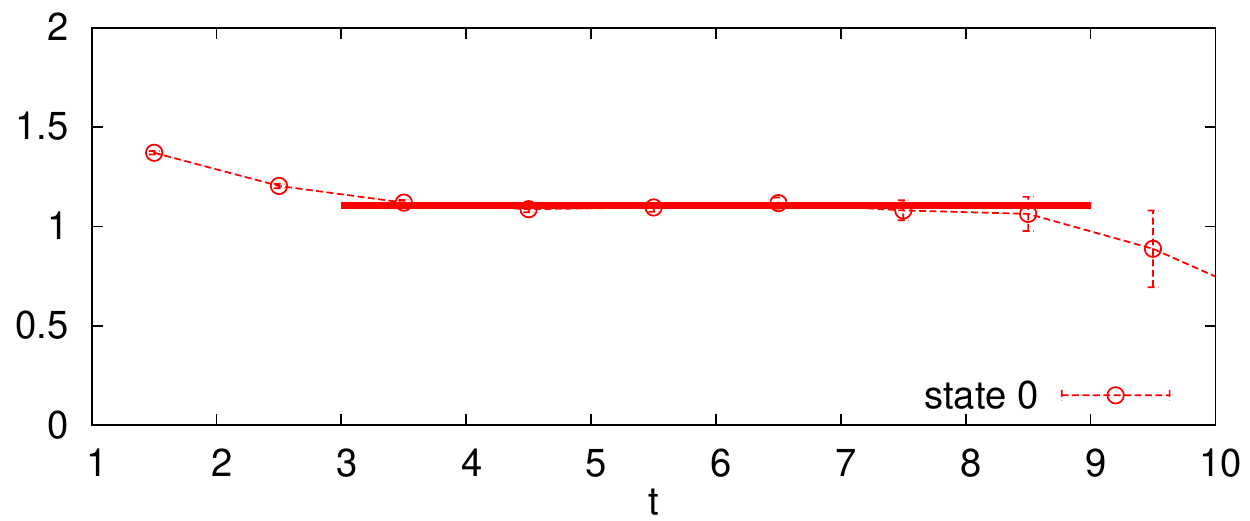}
  \hspace*{24pt}\hfil\\
  \caption{$b_1$ with 2 (upper row) and 128 (lower row) eigenmodes subtracted:
   The correlators for all eigenstates (left),
   effective mass plot for the lowest state (right).
}\label{fig:b1_1}
\end{figure*}

\section{Results and Discussion}\label{sec:results}

\subsection{Truncation study}

The reduction parameter $k$ in \eq{eq:red5} gives the number of the
lowest eigenmodes of the Dirac operator removed from the quark
propagator. We study reduced quark propagators with $k=0, 2, 4, 8, 12,
16, 20, 32, 64, 128$ and three different quark smearings ($n$, $w$,
$\partial_i$). These are then combined into different hadron propagators
and the correlation matrix for a hadron with given quantum numbers is
calculated. The variational method is used to extract  the ground and
excited states of that hadron. Consequently, we observe and study the
evolution of hadron masses as a function of the number of the subtracted
lowest eigenmodes. Increasing the number of the subtracted lowest eigenmodes
we gradually remove the chiral condensate of the vacuum and consequently
``unbreak'' the chiral symmetry.

\begin{figure*}
  \hspace*{24pt}
  \includegraphics[width=0.4\textwidth]{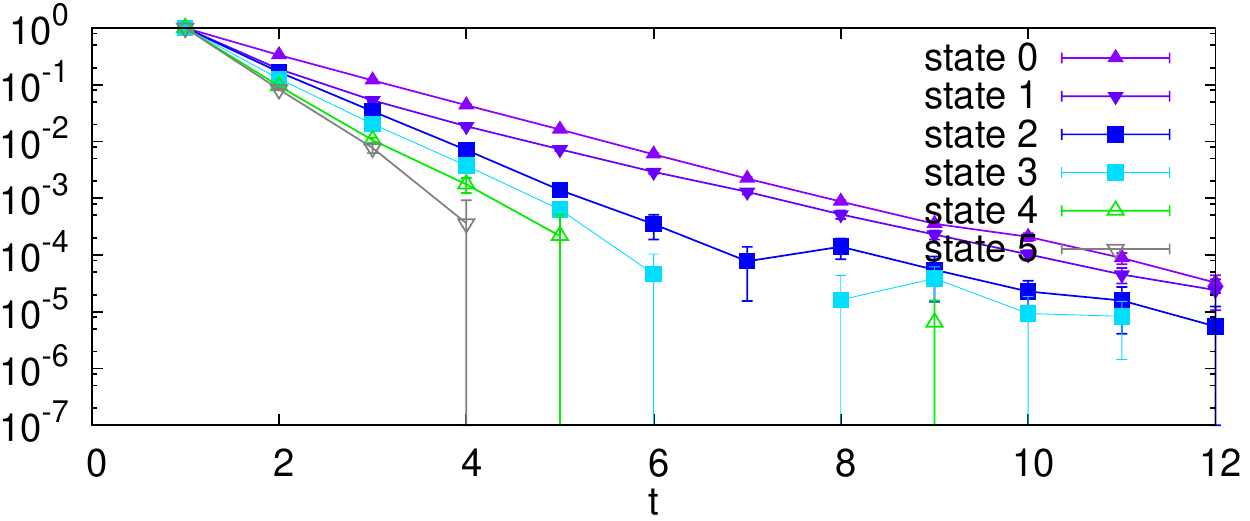}\hfill
  \includegraphics[width=0.4\textwidth]{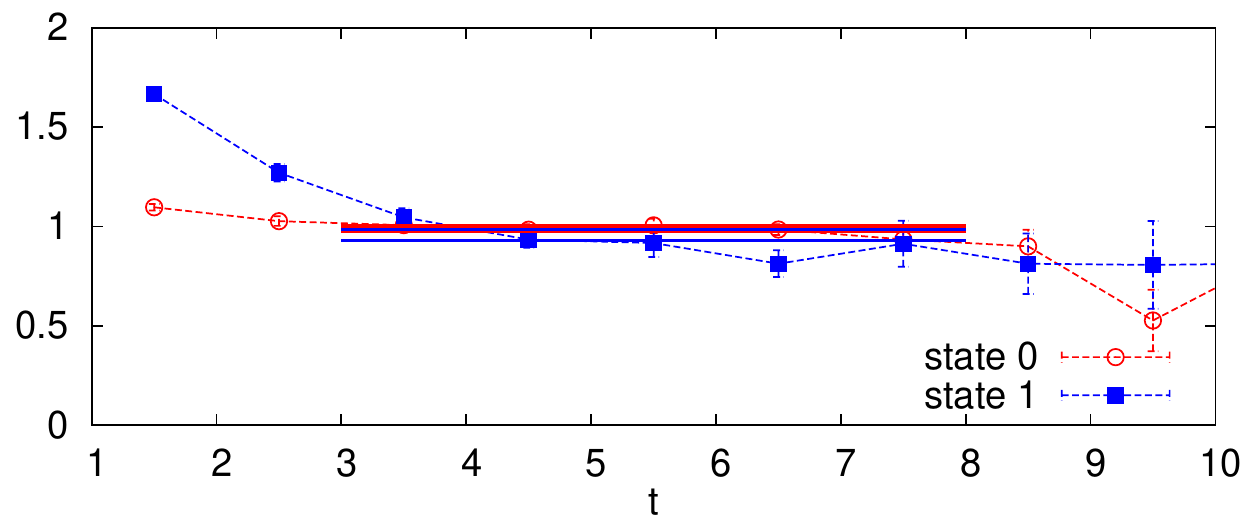}
  \hspace*{24pt}\hfil\\
  \hspace*{24pt}
  \includegraphics[width=0.4\textwidth]{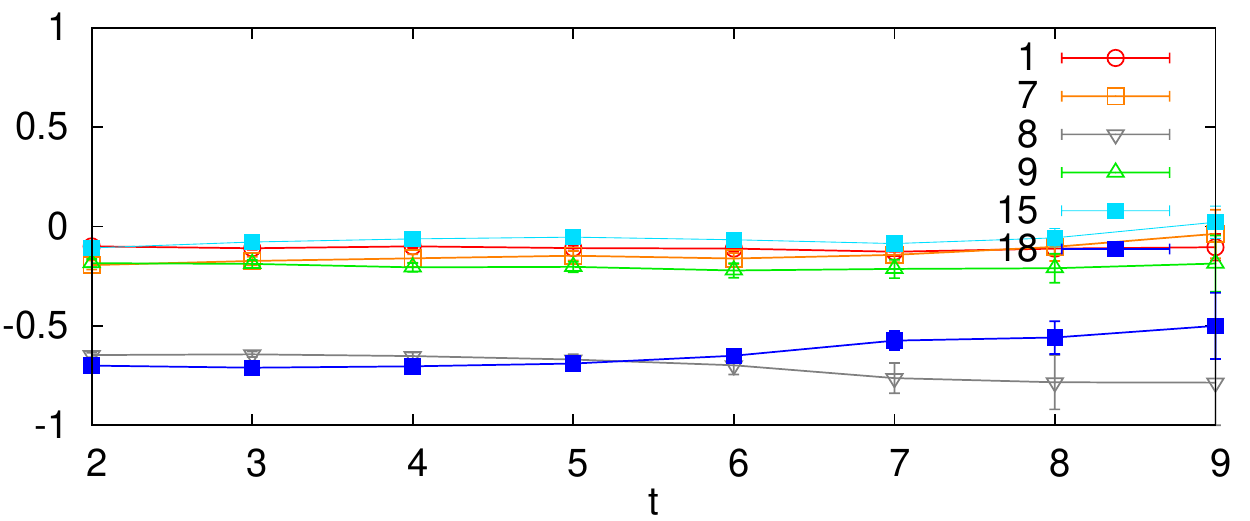}\hfill
  \includegraphics[width=0.4\textwidth]{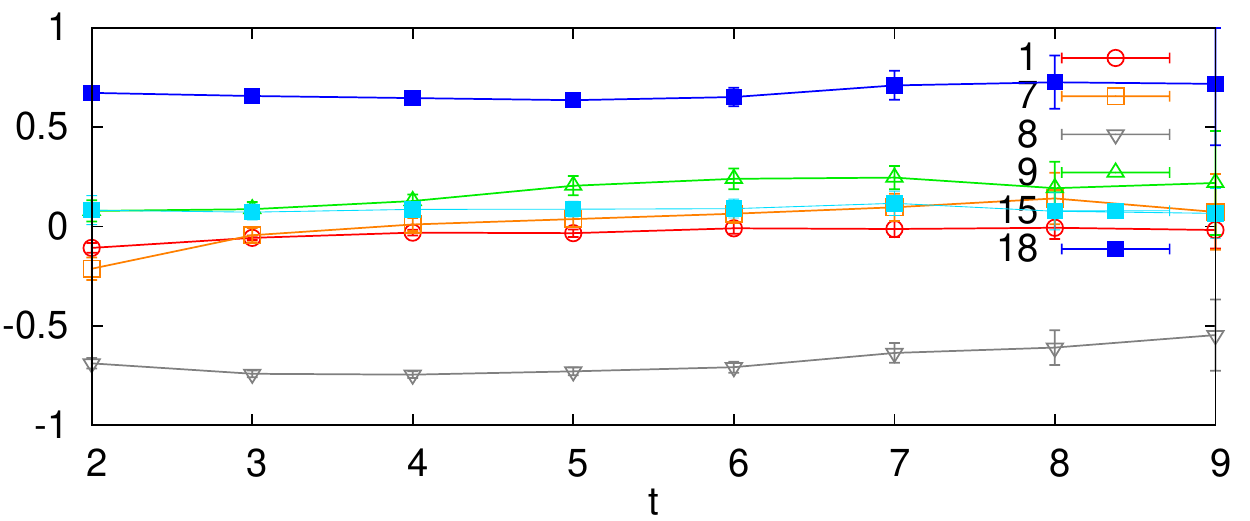}
  \hspace*{24pt}\hfil\\
  \caption{$N(+)$ with 20 eigenmodes subtracted: The correlators for all eigenstates (upper left),
   effective mass plot for the two lowest states (upper right),    
   eigenvectors corresponding to the ground state (lower left).
   and 1st excited state (lower right). 
}\label{fig:N_pos1}
\end{figure*}

\begin{figure*}
  \hspace*{24pt}
  \includegraphics[width=0.4\textwidth]{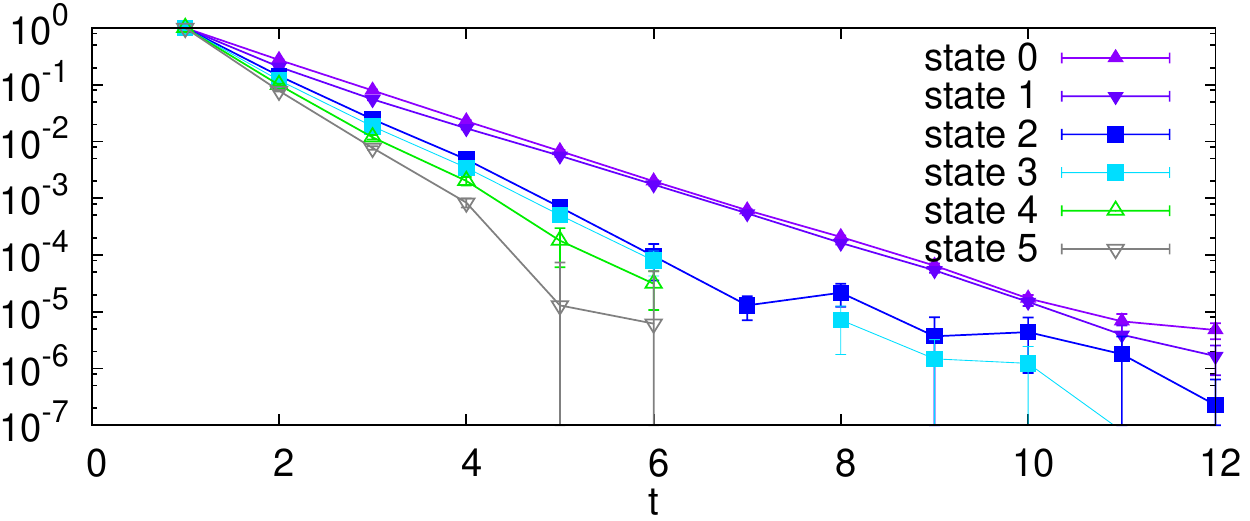}\hfill
  \includegraphics[width=0.4\textwidth]{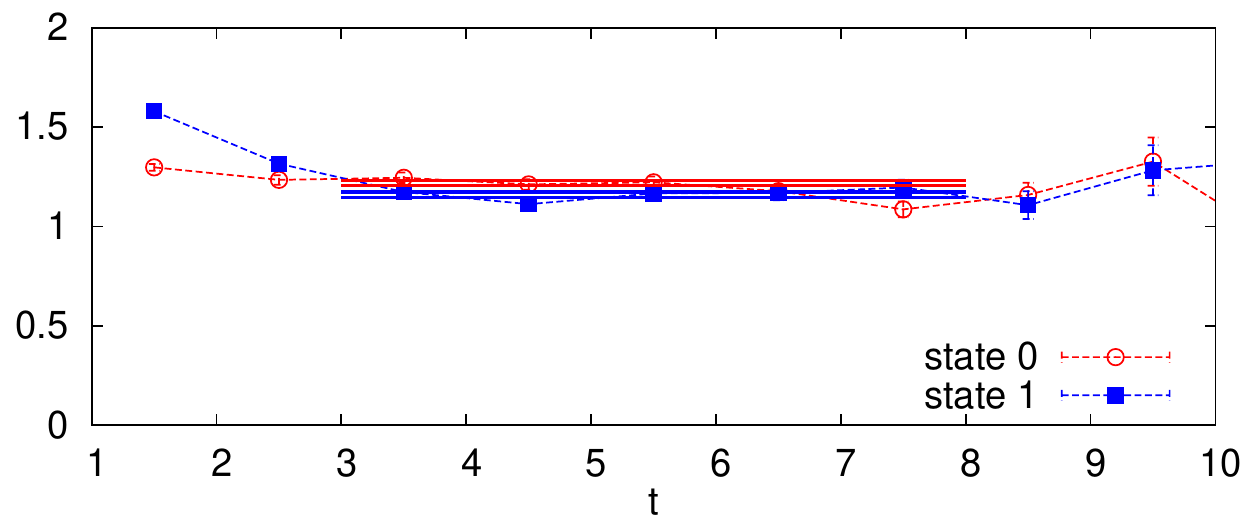}
  \hspace*{24pt}\hfil\\
  \hspace*{24pt}
  \includegraphics[width=0.4\textwidth]{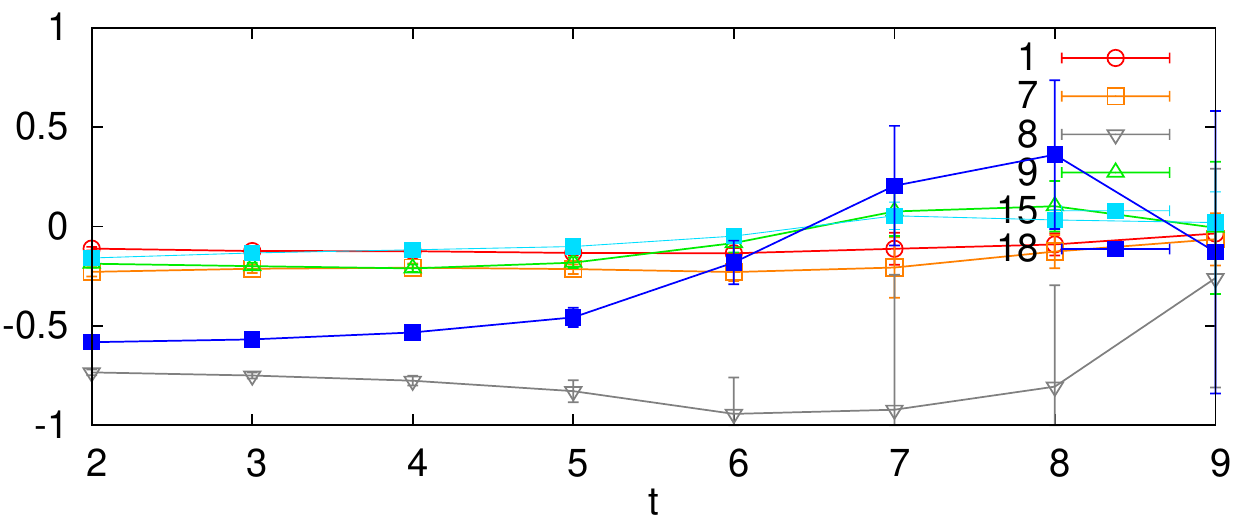}\hfill
  \includegraphics[width=0.4\textwidth]{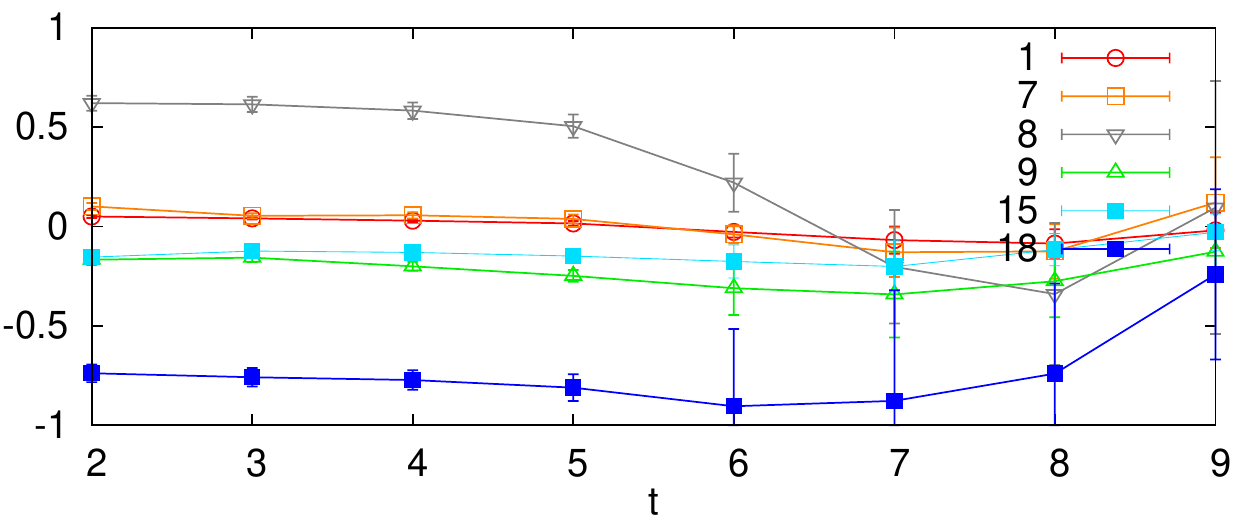}
  \hspace*{24pt}\hfil\\
  \caption{$N(+)$ with 64 eigenmodes subtracted: The correlators for all eigenstates (upper left),
   effective mass plot for the two lowest states (upper right),    
   eigenvectors corresponding to the ground state (lower left).
   and 1st excited state (lower right). 
}\label{fig:N_pos2}
\end{figure*}

\begin{figure*}
  \hspace*{24pt}
  \includegraphics[width=0.4\textwidth]{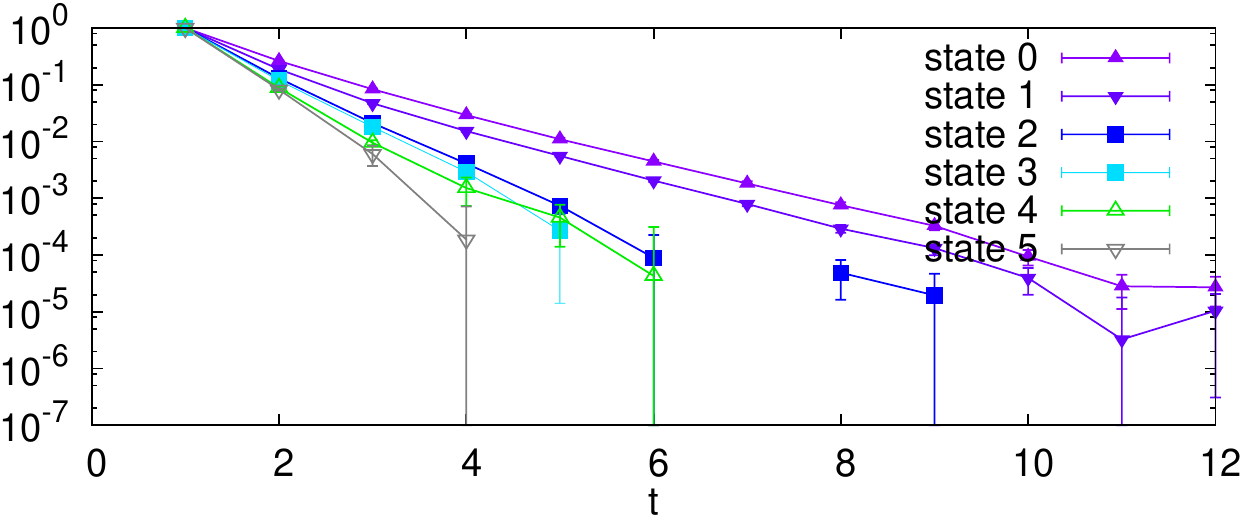}\hfill
  \includegraphics[width=0.4\textwidth]{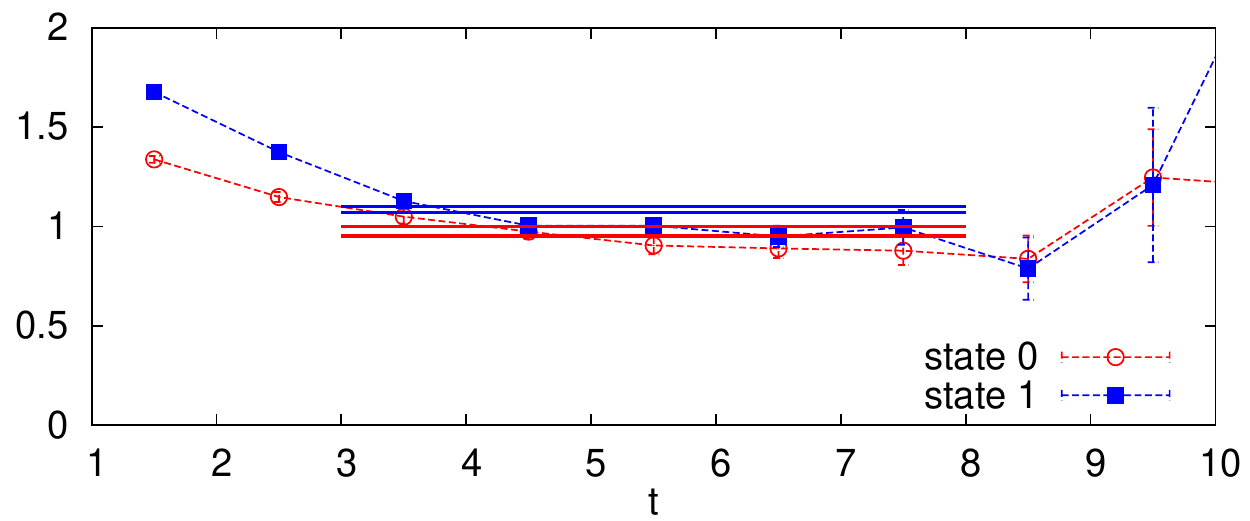}
  \hspace*{24pt}\hfil\\
  \hspace*{24pt}
  \includegraphics[width=0.4\textwidth]{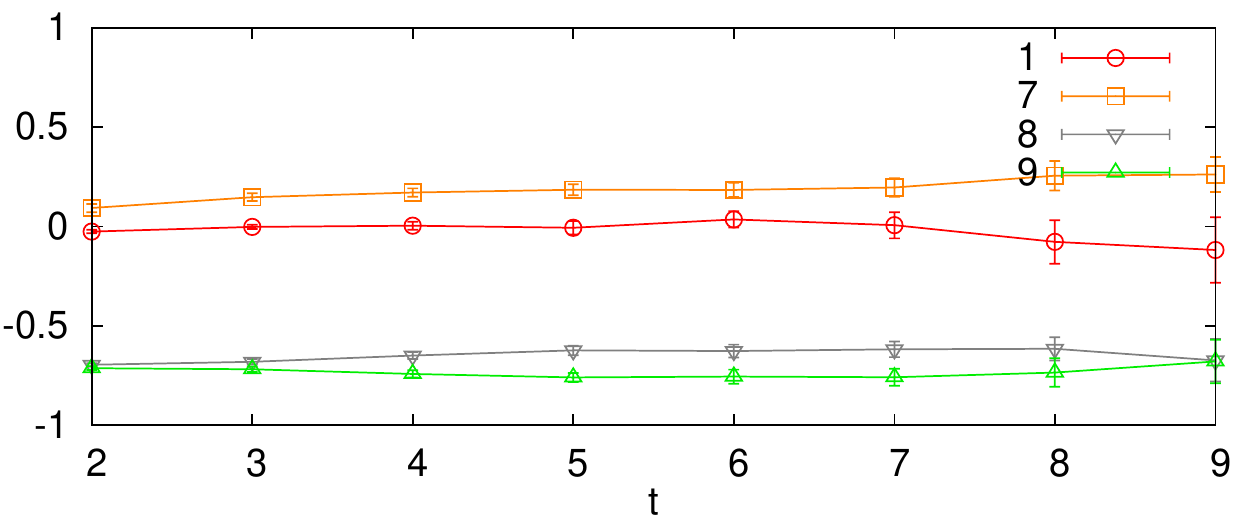}\hfill
  \includegraphics[width=0.4\textwidth]{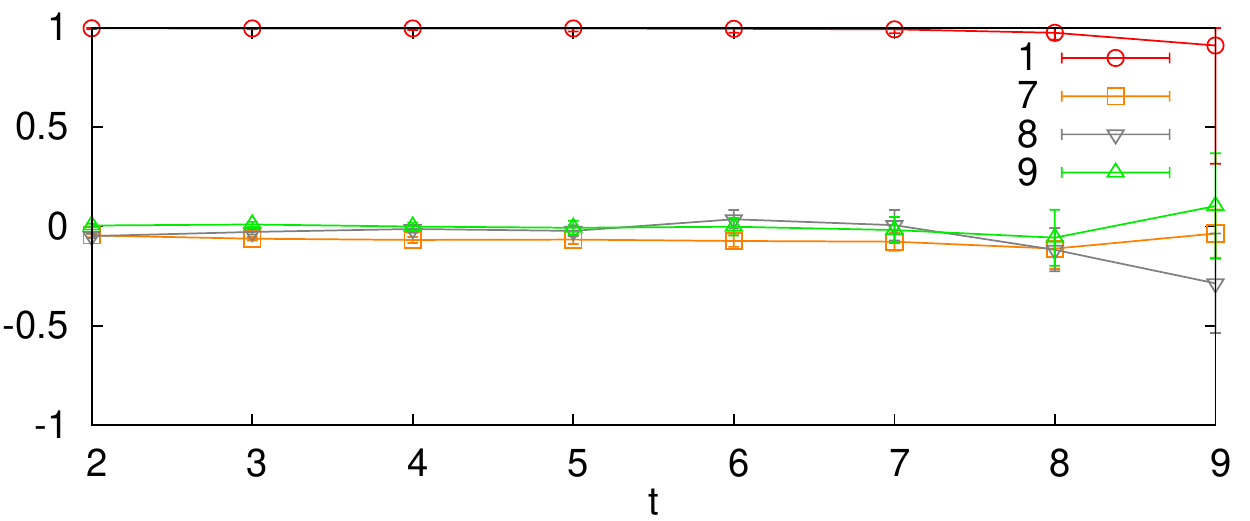}
  \hspace*{24pt}\hfil\\
  \caption{$N(-)$ with 12 eigenmodes subtracted: The correlators for all eigenstates (upper left),
   effective mass plot for the two lowest states (upper right),    
   eigenvectors corresponding to the ground state (lower left).
   and 1st excited state (lower right). 
}\label{fig:N_neg1}
\end{figure*}

\begin{figure*}
  \hspace*{24pt}
  \includegraphics[width=0.4\textwidth]{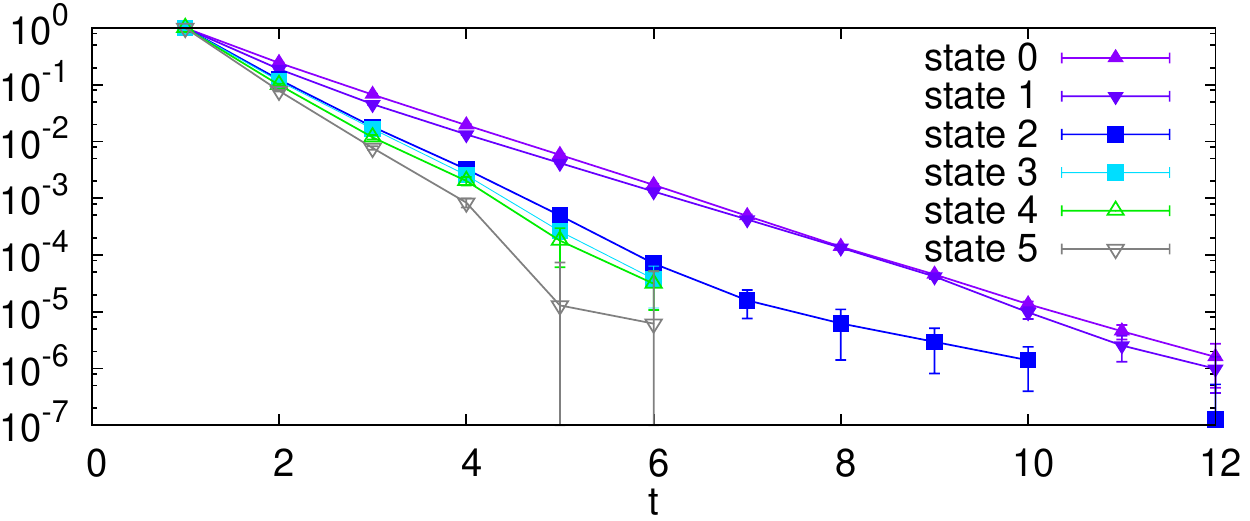}\hfill
  \includegraphics[width=0.4\textwidth]{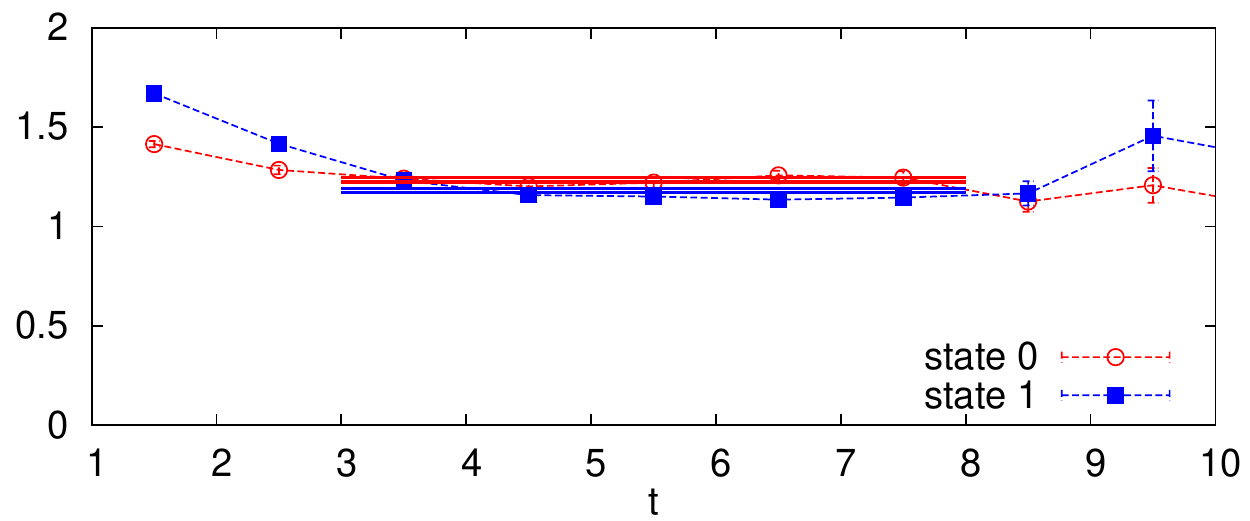}
  \hspace*{24pt}\hfil\\
  \hspace*{24pt}
  \includegraphics[width=0.4\textwidth]{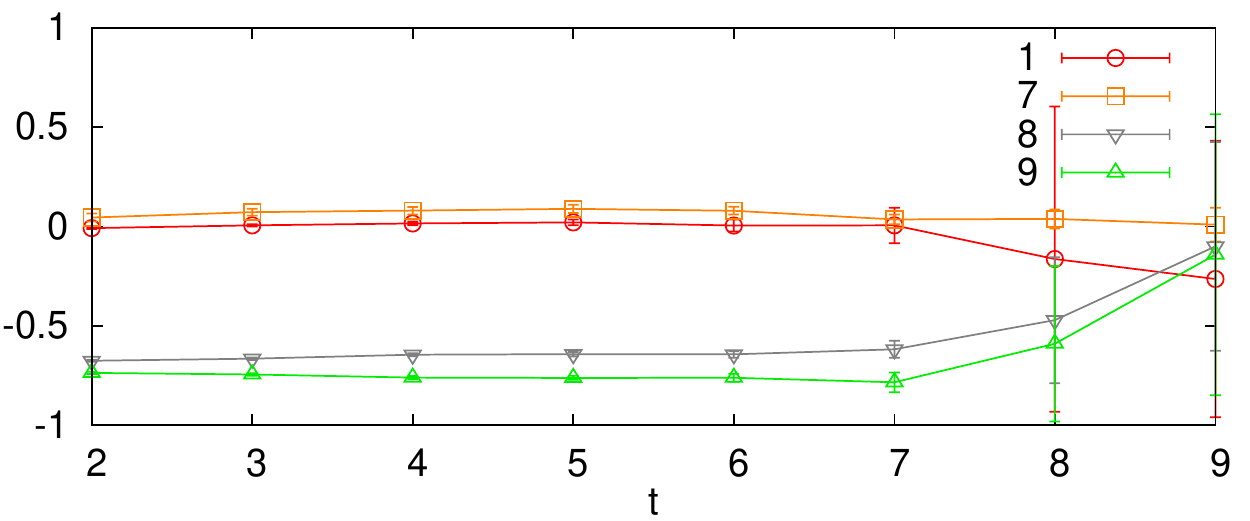}\hfill
  \includegraphics[width=0.4\textwidth]{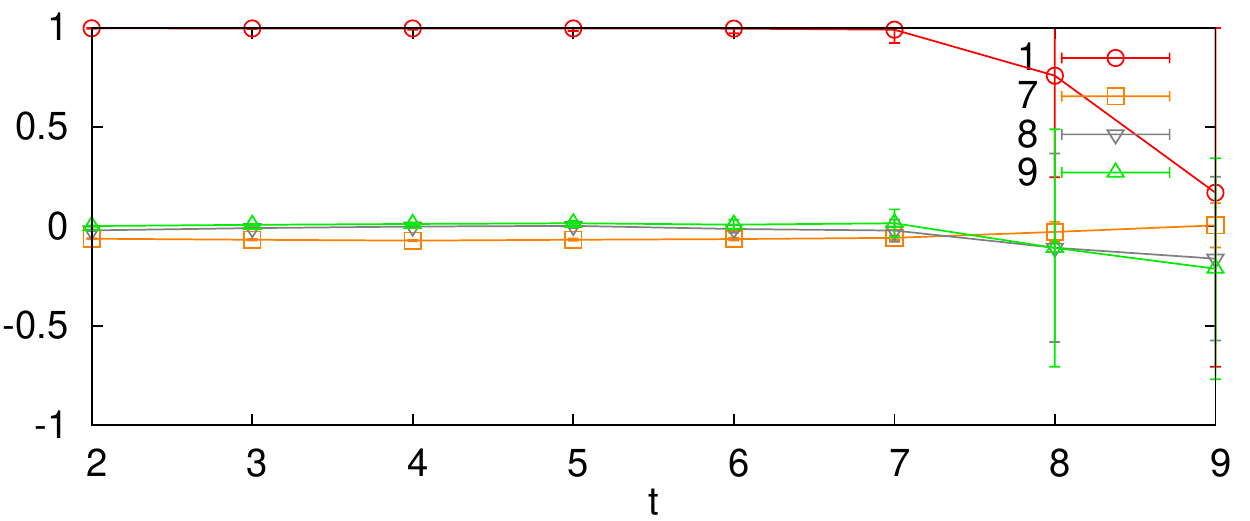}
  \hspace*{24pt}\hfil\\
  \caption{$N(-)$ with 64 eigenmodes subtracted: The correlators for all eigenstates (upper left),
   effective mass plot for the two lowest states (upper right),    
   eigenvectors corresponding to the ground state (lower left).
   and 1st excited state (lower right). 
}\label{fig:N_neg2}
\end{figure*}

\begin{figure*}
  \hspace*{24pt}
  \includegraphics[width=0.4\textwidth]{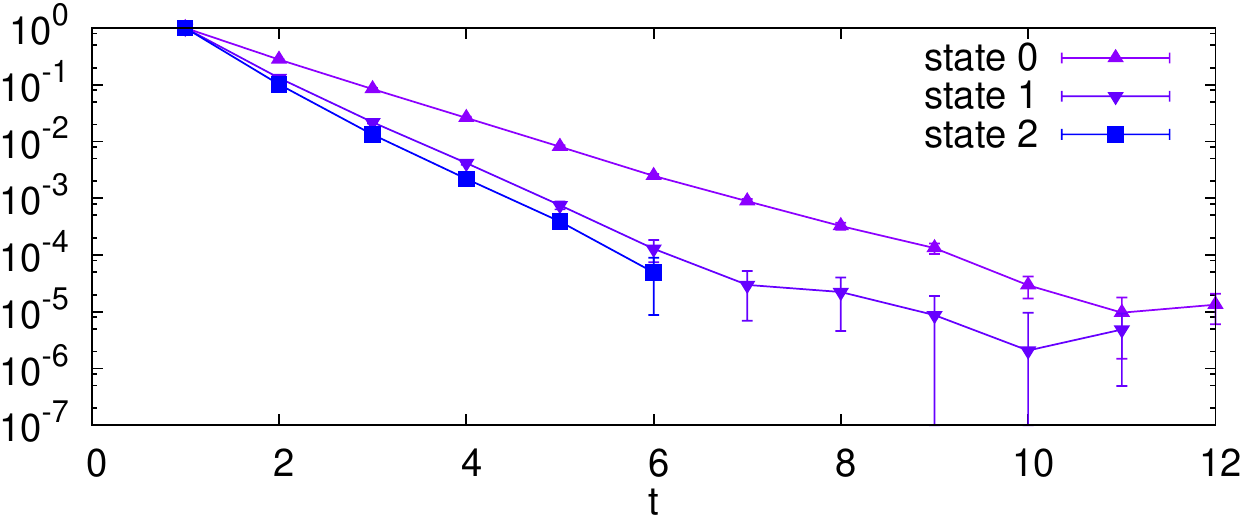}\hfill
  \includegraphics[width=0.4\textwidth]{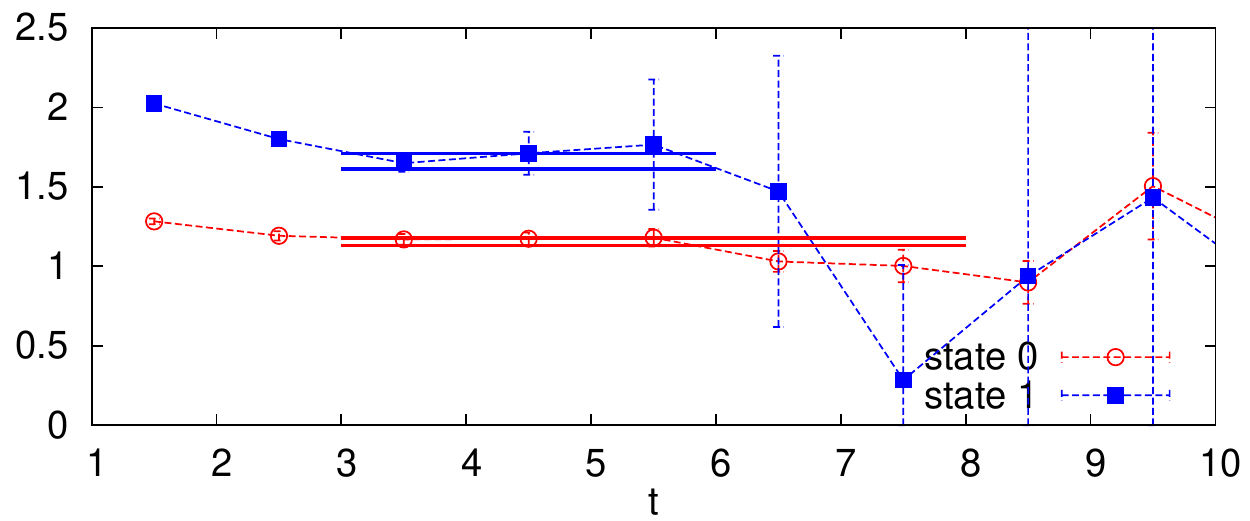}
  \hspace*{24pt}\hfil\\
  \hspace*{24pt}
  \includegraphics[width=0.4\textwidth]{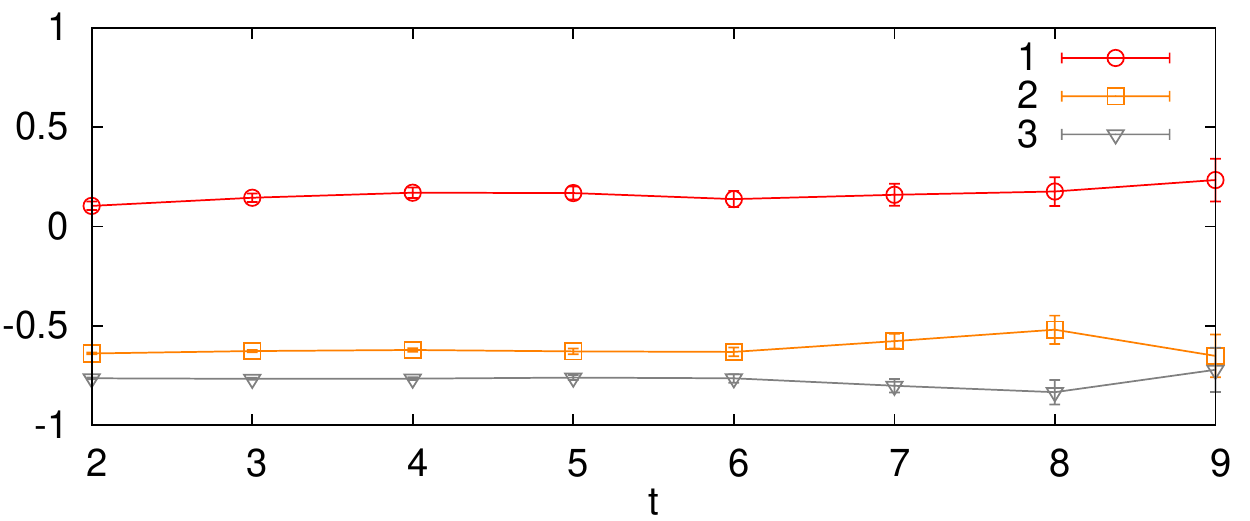}\hfill
  \includegraphics[width=0.4\textwidth]{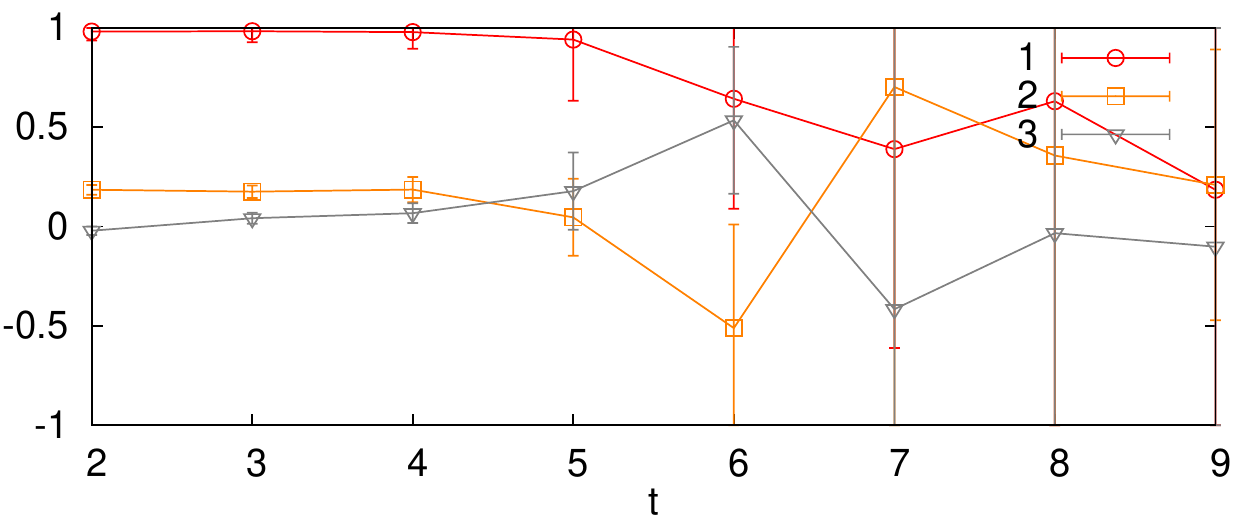}
  \hspace*{24pt}\hfil\\
  \caption{$\Delta(+)$ with 16 eigenmodes subtracted: The correlators for all eigenstates (upper left),
   effective mass plot for the two lowest states (upper right),    
   eigenvectors corresponding to the ground state (lower left).
   and 1st excited state (lower right). 
}\label{fig:Delta_pos1}
\end{figure*}

\begin{figure*}
  \hspace*{24pt}
  \includegraphics[width=0.4\textwidth]{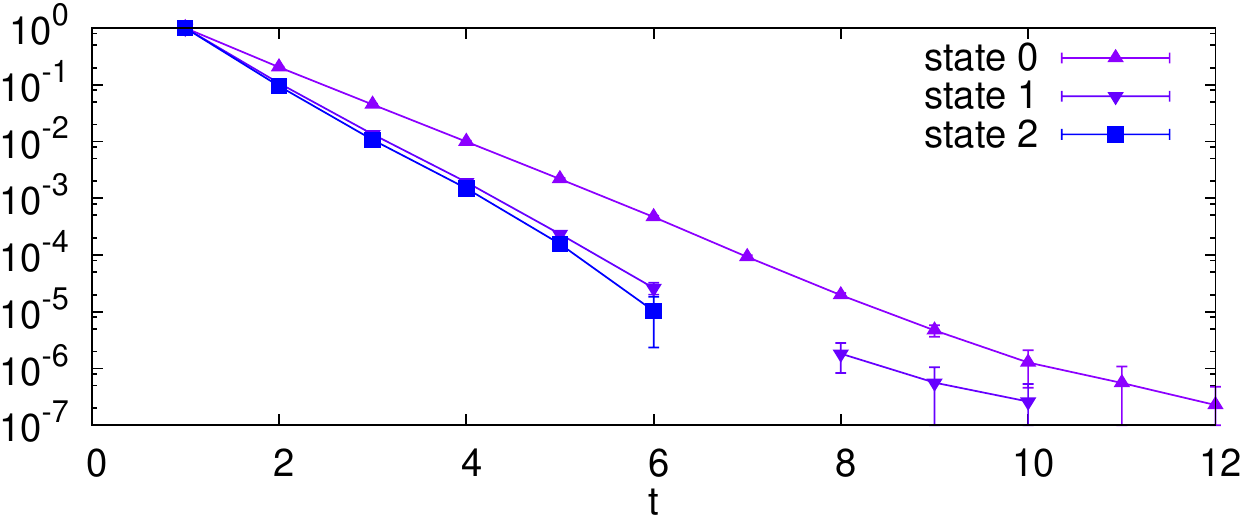}\hfill
  \includegraphics[width=0.4\textwidth]{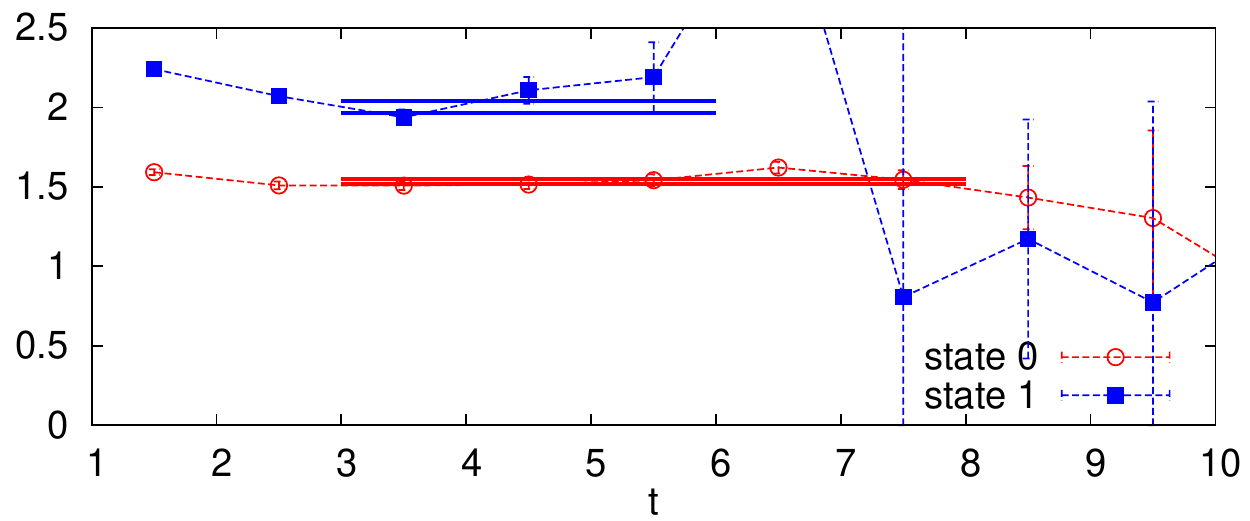}
  \hspace*{24pt}\hfil\\
  \hspace*{24pt}
  \includegraphics[width=0.4\textwidth]{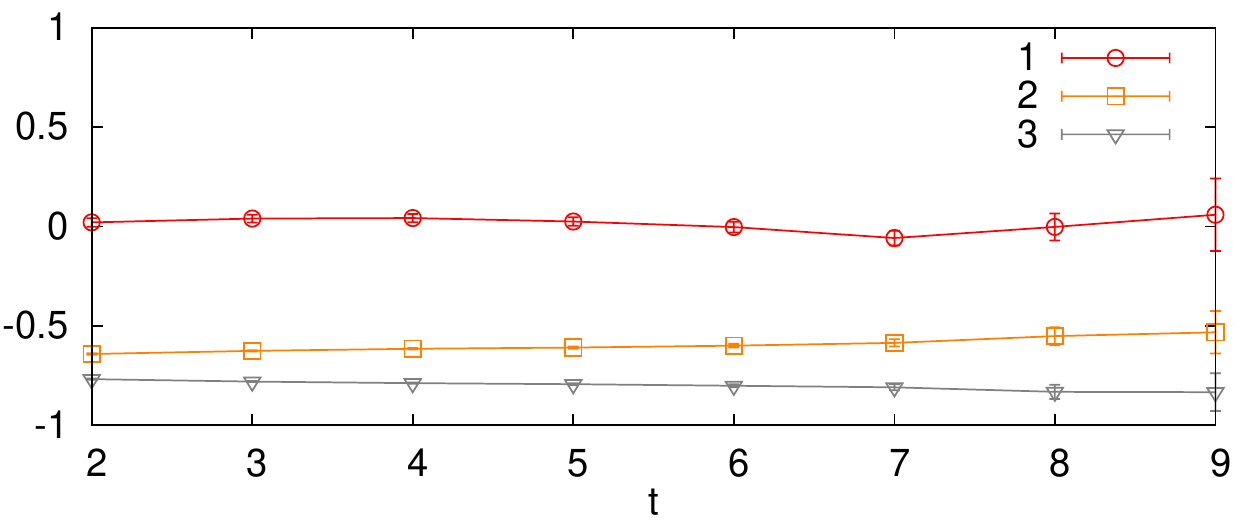}\hfill
  \includegraphics[width=0.4\textwidth]{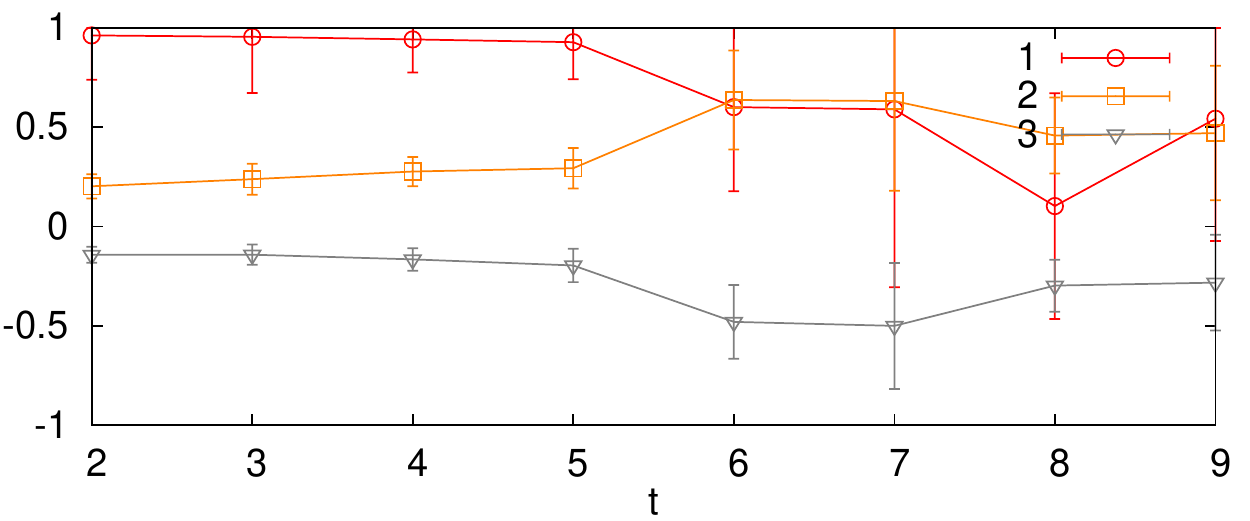}
  \hspace*{24pt}\hfil\\
  \caption{$\Delta(+)$ with 128 eigenmodes subtracted: The correlators for all eigenstates (upper left),
   effective mass plot for the two lowest states (upper right),    
   eigenvectors corresponding to the ground state (lower left).
   and 1st excited state (lower right). 
}\label{fig:Delta_pos2}
\end{figure*}

\begin{figure*}
  \hspace*{24pt}
  \includegraphics[width=0.4\textwidth]{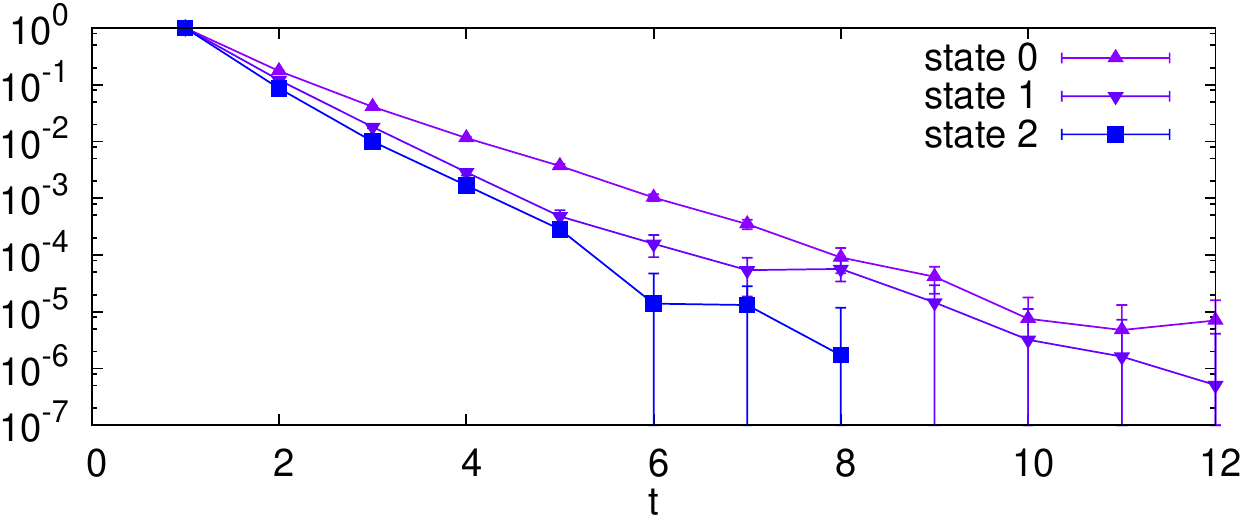}\hfill
  \includegraphics[width=0.4\textwidth]{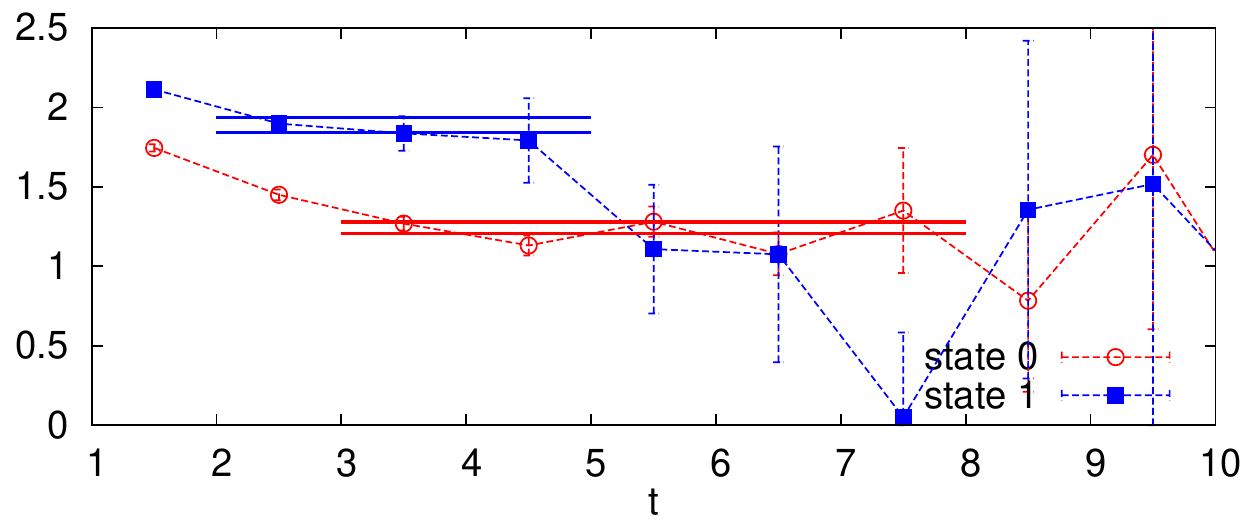}
  \hspace*{24pt}\hfil\\
  \hspace*{24pt}
  \includegraphics[width=0.4\textwidth]{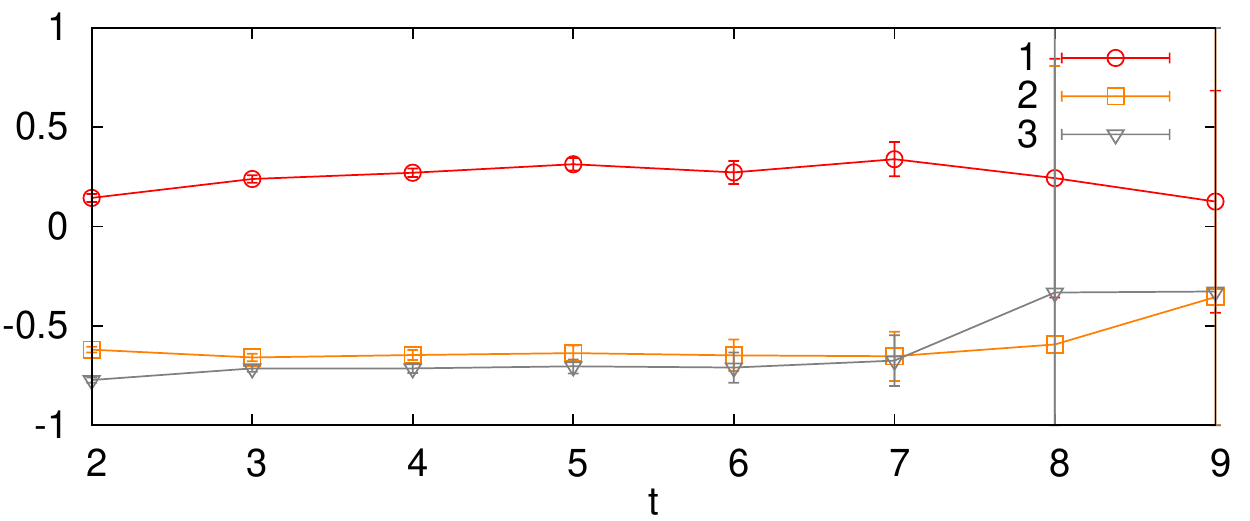}\hfill
  \includegraphics[width=0.4\textwidth]{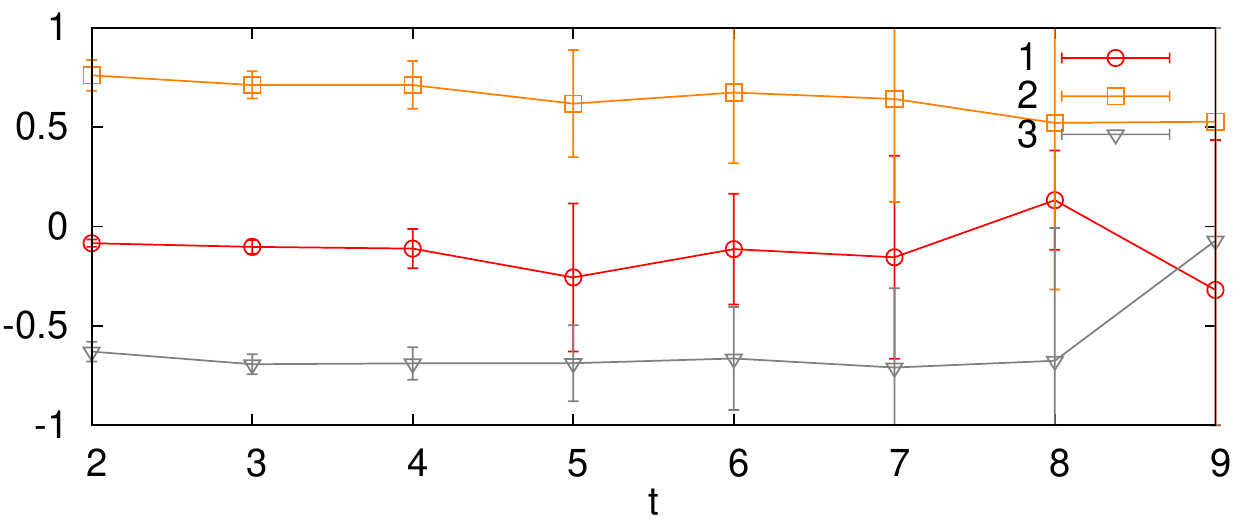}
  \hspace*{24pt}\hfil\\
  \caption{$\Delta(-)$ with 16 eigenmodes subtracted: The correlators for all eigenstates (upper left),
   effective mass plot for the two lowest states (upper right),    
   eigenvectors corresponding to the ground state (lower left).
   and 1st excited state (lower right). 
}\label{fig:Delta_neg1}
\end{figure*}

\begin{figure*}
  \hspace*{24pt}
  \includegraphics[width=0.4\textwidth]{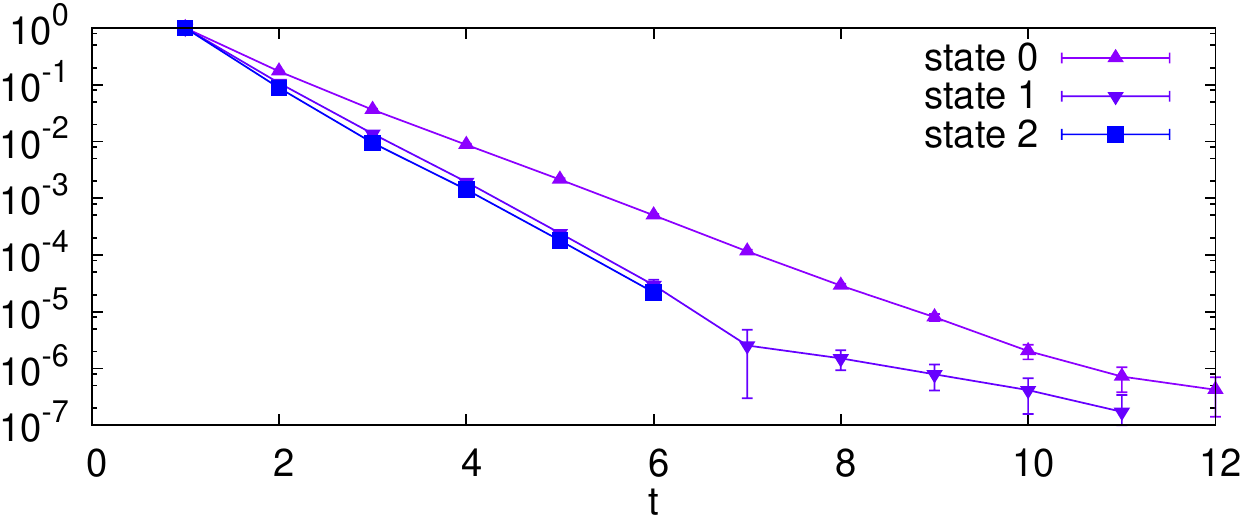}\hfill
  \includegraphics[width=0.4\textwidth]{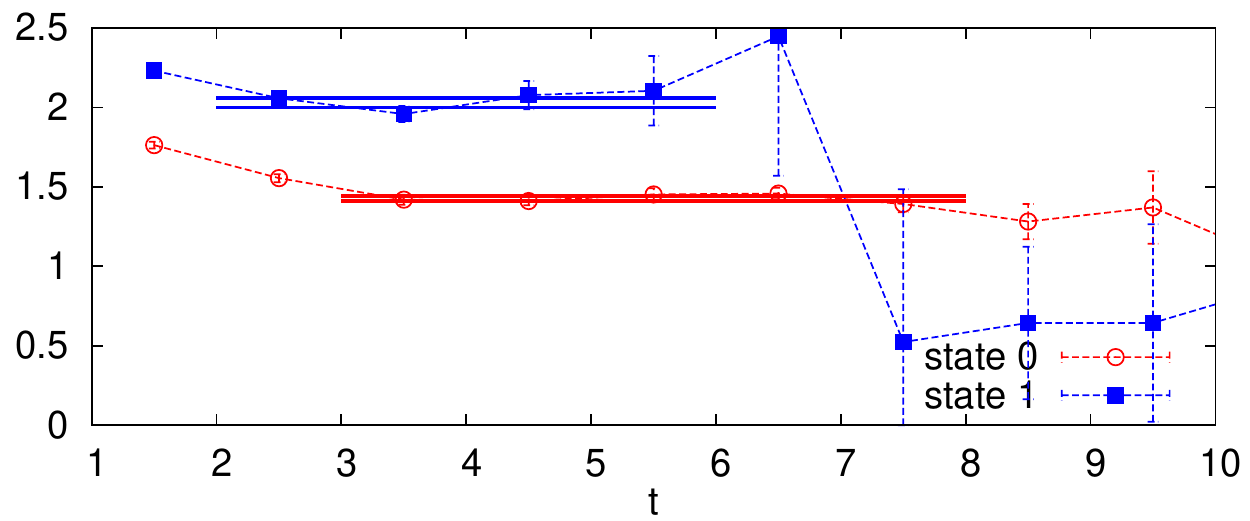}
  \hspace*{24pt}\hfil\\
  \hspace*{24pt}
  \includegraphics[width=0.4\textwidth]{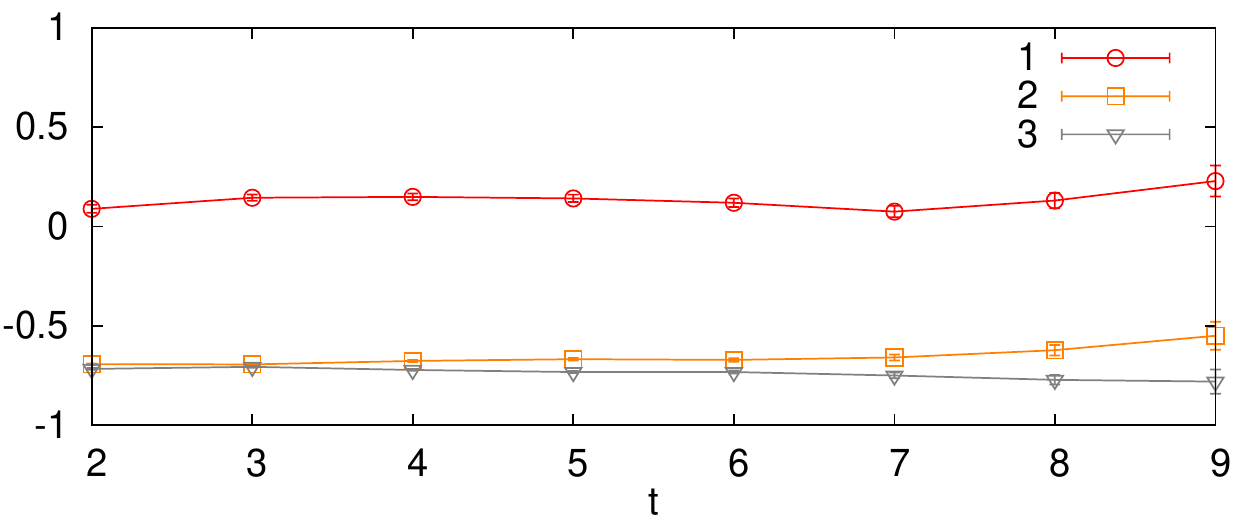}\hfill
  \includegraphics[width=0.4\textwidth]{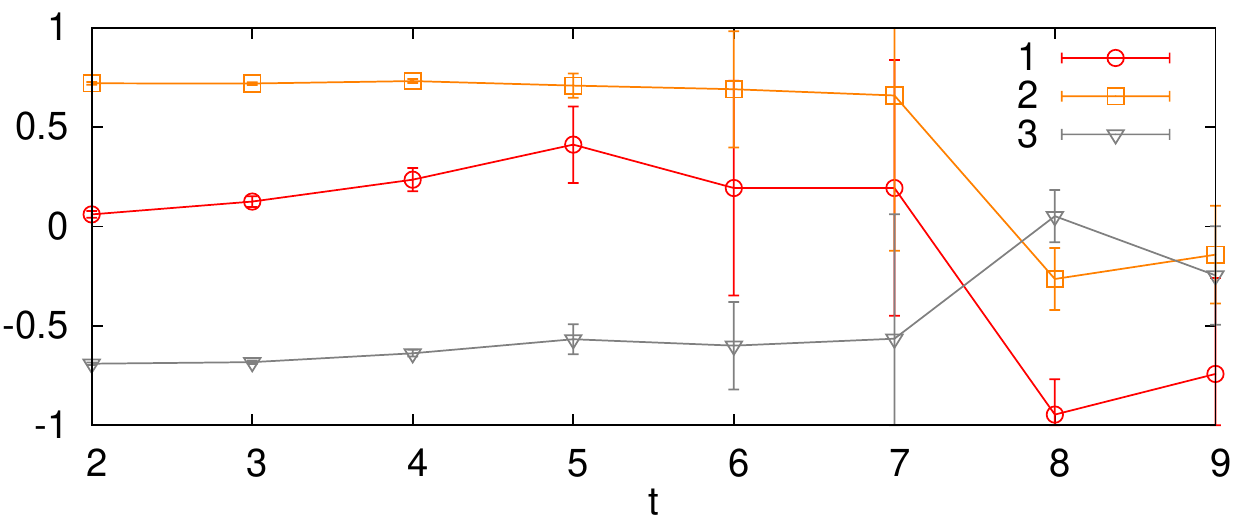}
  \hspace*{24pt}\hfil\\
  \caption{$\Delta(-)$ with 128 eigenmodes subtracted: The correlators for all eigenstates (upper left),
   effective mass plot for the two lowest states (upper right),    
   eigenvectors corresponding to the ground state (lower left).
   and 1st excited state (lower right). 
}\label{fig:Delta_neg2}
\end{figure*}

Typical results for the mesons and baryons under study are shown in 
Figs.~\ref{fig:rho_1} - \ref{fig:Delta_neg2}. For each hadron we show in
the figures two representative reduction levels $k$ for which we also
show explicitly all eigenvalues stemming from the variational  method,
i.e., the correlators corresponding to different energy levels. Moreover,
we show for the ground and first excited state (where applicable) the
eigenvector components and the effective mass plots including  fit ranges
and values. The energy values are determined from exponential fits  to
the eigenvalues over the indicated fit ranges.

In Figs. \ref{fig:rho_1} and \ref{fig:rho_2} we show the eigenvalues and
eigenvectors (cf. Sect. \ref{sec:varmeth})
and the effective mass plots for the ground and excited
states of the $\rho$-meson ($J^{PC}=1^{--}$) after having subtracted 12
and 32 eigenmodes of $D_5$. (The results  for the untruncated situation,
with higher statistics and for several more parameter sets are shown in
\cite{Engel:2011aa}.) The eigenvector composition for both states is
stable and clearly distinct. The mass splitting between ground state and
excited state disappears with increasing truncation level. 

For the meson channels $a_1$ and $b_1$ the statistics allow only to
determine the ground state in a reliable way. In the case of the $a_1$
($J^{PC}=1^{++}$) meson  (\fig{fig:a1_1} for $k=4$ and $k=64$) and for
the $b_1$ ($J^{PC}=1^{+-}$) meson (\fig{fig:b1_1} for $k=2$ and $k=128$)
we thus show only the ground states. We observe improved plateau quality
for the effective  masses when  increasing number of truncated modes.

In \fig{fig:N_pos1}  the nucleon of positive parity (the ground state and
its first excitation, $J^P= \ot^+$), after having subtracted the lowest
20 eigenmodes is shown. These should be compared with  \fig{fig:N_pos2}
where 64 modes have been  subtracted. In \fig{fig:N_neg1} and
\ref{fig:N_neg2} we present two nucleon states of negative parity, $J^P=
\ot^-$, (at reduction level 12 and 64).

The positive parity $\Delta$  ground and excited states, $J^P=
\frac{3}{2}^+$ at reduction level 16 and 128 are shown in
\fig{fig:Delta_pos1} and \fig{fig:Delta_pos2} and the  negative parity
$\Delta$'s ($J^P= \frac{3}{2}^-$)  at the same reduction levels are given
in  \fig{fig:Delta_neg1} and \fig{fig:Delta_neg2}, respectively.

An obvious observation is that the quality of the signal (the quality of
plateaus) essentially improves with increasing the number of removed
eigenmodes for all hadrons under study. This fact makes it easier to
reliably identify masses of states after unbreaking the chiral symmetry.
In many cases for excited hadrons the quality of  plateaus before
unbreaking of the chiral symmetry is rather poor, but after removing
more and more eigenmodes of the Dirac operator it becomes better and
better, so eventually it allows to unambiguously establish that indeed we
see the state, even though with the untruncated propagators the
identification of the state would be less clear. 

A plausible explanation for this phenomenon would be that by unbreaking
the chiral symmetry we remove from the hadron its pion cloud and subtract
all higher Fock components like $\pi N$, $\pi \Delta$, $\pi \pi$, and so
on from the hadron wave functions. It is these components related to the
chiral symmetry breaking  that couple excited and not excited hadrons to
each other and render signals from the excited states poor in fully
untruncated QCD.

\subsection{Confinement after unbreaking the chiral symmetry}

The most interesting question is whether hadrons and confinement survive
the unbreaking of the chiral symmetry.  To discuss this issue we put all
the results from the previous figures  together and analyze how  masses
of the considered  hadrons change with  the reduction level. Therefore,
the relevant  truncation scale is not the Dirac operator eigenmode index
$k$ itself, since it has to scale with the lattice volume when keeping
the  physics constant. Instead, we introduce a cutoff parameter $\sigma$
such that the reduction level $\sigma$ means that all $\mu_k$ for which
$|\mu_k|<\sigma$  have been excluded in the underlying quark propagators
\cite{Lang:2011qy}. We still give the corresponding index $k$ on the
upper horizontal scale of the plots.

The masses of all studied hadrons under $D_5$ eigenmode reduction are
summarized in \fig{fig:summary}. The scale  is set by the Sommer
parameter (i.e., by the static potential acting between two heavy quarks,
in other words, by the gluonic dynamics).  The physics of the static
potential knows nothing about the valence quarks and the truncation of
lowest lying eigenmodes of the Dirac operator for the light valence
quarks. It implies that these plots suggest at least qualitatively the
evolution of the hadron masses in absolute units of energy. We observe
approximately a universal growth of all hadron masses with  equal slope
after subtraction of a sufficient amount of the chiral modes of the Dirac
operator. In this regime chiral symmetry is approximately restored and
the whole hadron mass is not related to chiral symmetry breaking. A
universal slope might be  interpreted as an  indication to a universal
growth of the hadron size. 

Assume that after having unbroken the chiral symmetry the exponential decay
signals from all  hadrons would disappear. This would indicate that with
the artificial restoration of the chiral symmetry confinement also
vanishes and that there is a direct connection between the confinement in
QCD and the  lowest lying modes of the Dirac operator. Contrary to that
we observe a very clear signal from all hadrons, except for a pion. This
suggests that confinement survives the unbreaking of the chiral symmetry.

However, there is still the possibility that this clear signal comes from
the unconfined (unbound) quarks with some mass $m_0$ at a given truncation level.
In this case we would expect at a given truncation level a universal
scaling law  with all mesons having the mass $2\,m_0$ and all baryons with
the mass $3\,m_0$. There would be no excited states of hadrons.

In order to address this issue we  show in \fig{fig:summary_ratios} all 
hadron masses in units of the $\rho$-meson mass obtained at the same
truncation level. Indeed, some of the states -- such as $a_1$ and $\rho'$, as
well as the ground and the first excited states of the nucleon of both
parities -- do follow this behavior of mass $2\,m_0$ and $3\,m_0$
for mesons and baryons, respectively. However, this is 
definitely not the case for the $b_1$ state as well as for the
$\Delta$-resonance and especially its first excited states of positive
and negative parity. Given that the signal in all latter cases is
unambiguous, we conclude that there is no universal scaling ($2\,m_0$ for
mesons and  $3\,m_0$ for baryons) for all hadrons. This rules out the
possibility that our signals  are produced by the unbound (unconfined) 
quarks. We do observe confined hadrons.

Actually, this can be seen also from another perspective.  The mass
$m_0$ is large and increases with truncation of the quark propagators.
At the same time we observe chiral restoration in the correlators (e.g.,
$a_1$ and $\rho$). The large mass $m_0$ of unconfined free quarks then
contradicts restoration of chiral symmetry. This supports our argument
that we do not observe unconfined quarks.

The fact that masses of some of the mesons and some of the baryons get
degenerate and are related through a simple law   $2\,m_0$ for mesons and 
$3\,m_0$ for baryons indicates symmetries of hadrons that appear after
unbreaking (restoration) of the chiral symmetry.

\subsection{Meson degeneracies and splittings and what they tell us}

Restoration of the $SU(2)_L \times SU(2)_R$ chiral symmetry in the vacuum
requires  the states to fall into parity-chiral multiplets
\cite{Cohen:2001gb,Cohen:2002st,
Glozman:2002cp,Glozman:2003bt,Glozman:2007ek}. Below we shortly summarize
a content of these parity-chiral multiplets for the mesons in our study.

The full set of multiplets of the parity-chiral group $SU(2)_L \times
SU(2)_R\times C_i$, where the group $C_i$ consists of identity and the
space inversion,  for the $J=1$ mesons is as follows:
\begin{center}
\begin{tabular}{lclcl}
$(0,0)      $&$:\quad$&$\omega(0,1^{--}) $&$\qquad$ &$f_1(0,1^{++}) $ \nonumber\vspace{6pt}\\
$(\otf,\otf)_a$&$:\quad$&$h_1(0,1^{+-})    $&$\qquad$ &$\rho(1,1^{--})$ \nonumber\vspace{6pt}\\
$(\otf,\otf)_b$&$:\quad$&$\omega (0,1^{--})$&$\qquad$ &$b_1(1,1^{+-}) $ \nonumber\vspace{6pt}\\
$(0,1)+(1,0)$&$:\quad$&$a_1(1,1^{++})    $&$\qquad$ &$\rho (1,1^{--}$)\nonumber
\end{tabular}
\end{center}
Note, that the unbroken chiral $SU(2)_L \times SU(2)_R$ symmetry requires
existence of two independent $\rho$-mesons, one of them is the chiral
partner of the $h_1$ meson, and the other one of the $a_1$ state. Similar
is true for the $\omega$-meson. 

The states from two distinct multiplets $(\otf,\otf)_a$ and 
$(\otf,\otf)_b$ that have the same isospin but opposite spatial parity
are connected to each other by the $U(1)_A$ transformation, if the
$U(1)_A$ symmetry is broken neither explicitly nor spontaneously. In our
real world $U(1)_A$ is broken both explicitly via the axial anomaly and
spontaneously via the quark condensate of the vacuum. So in the world
with restored $U(1)_A$ symmetry a $\rho$ meson, that is the chiral
partner to the $h_1$ meson, would be degenerate with the $b_1$ state. The
$h_1,\,\rho,\,\omega$ and $b_1$ states would form an irreducible multiplet of
the $SU(2)_L \times SU(2)_R \times U(1)_A$ group.

\begin{figure*}
	\includegraphics[width=0.48\textwidth]{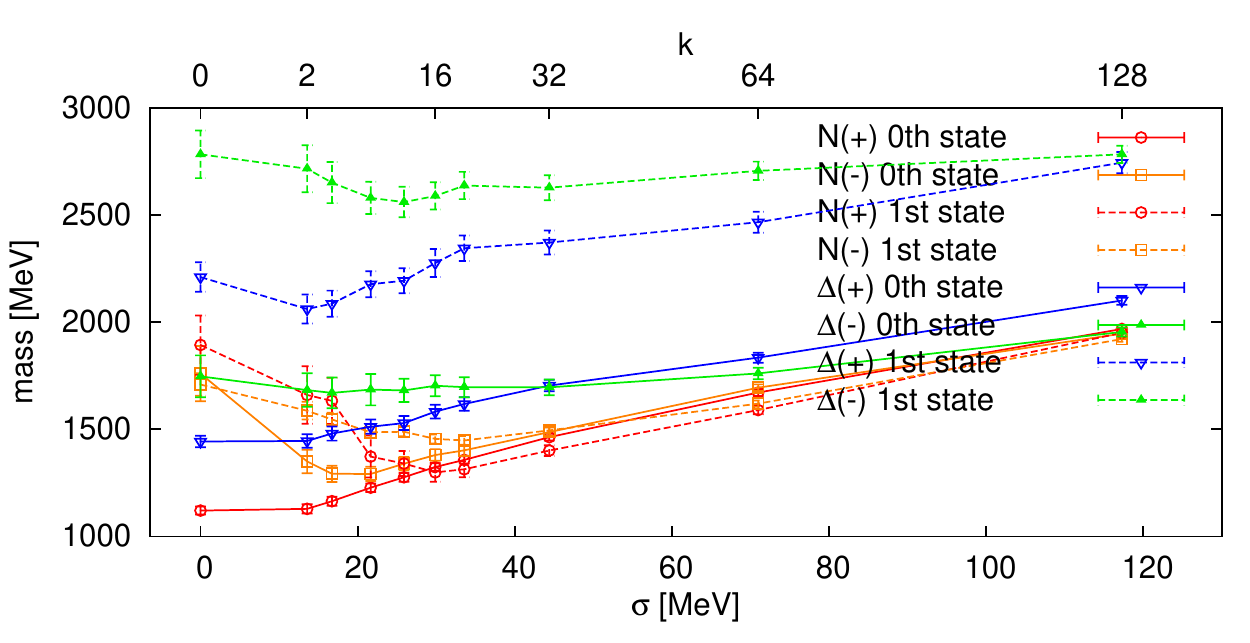}
	\hfill
        \includegraphics[width=0.48\textwidth]{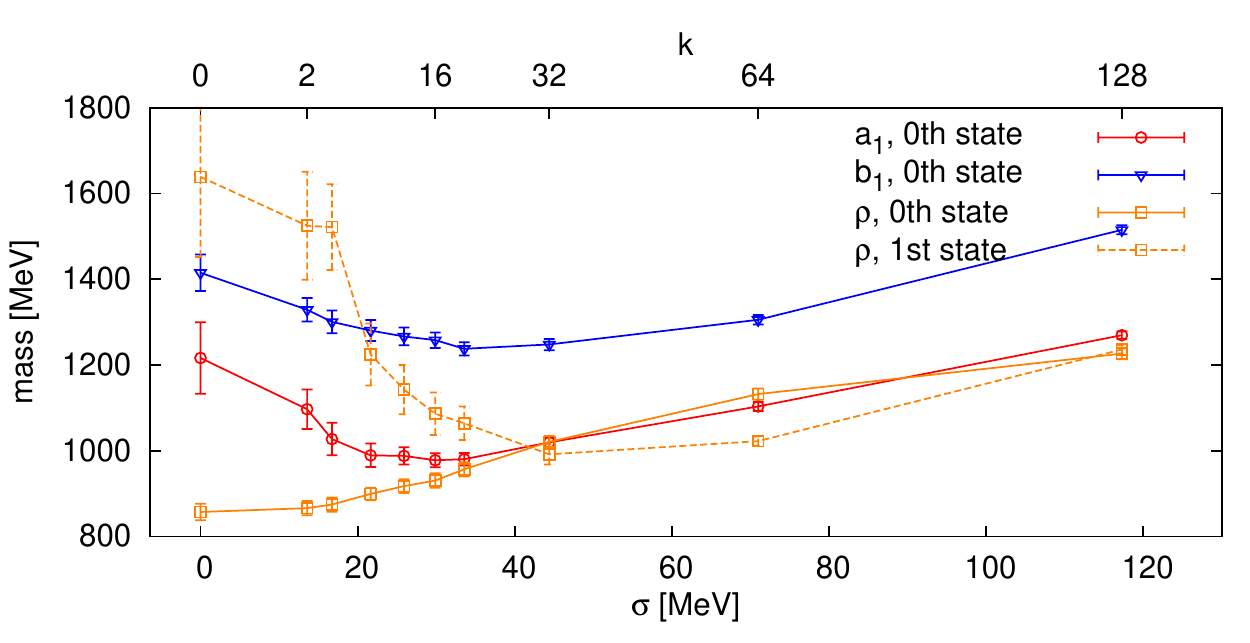}
	\caption{Summary plots: Baryon (l.h.s.) and meson (r.h.s.) masses
		as a function of the truncation level.}
        \label{fig:summary}
\end{figure*}

\begin{figure*}
	\includegraphics[width=0.48\textwidth]{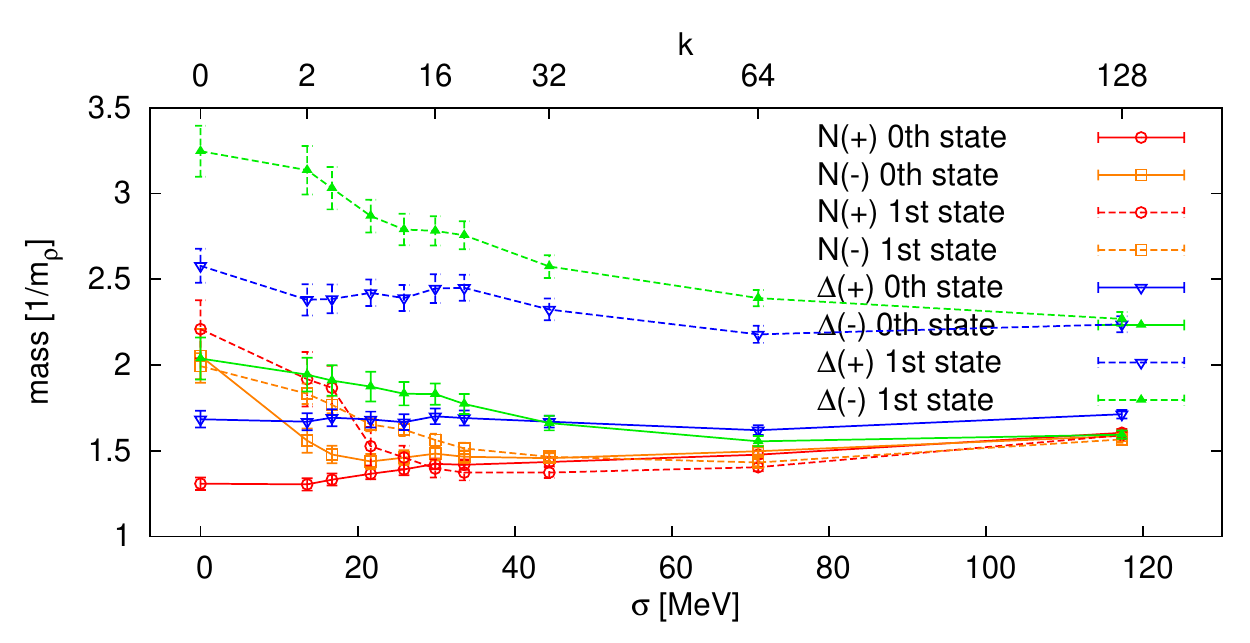}
	\hfill
	\includegraphics[width=0.48\textwidth]{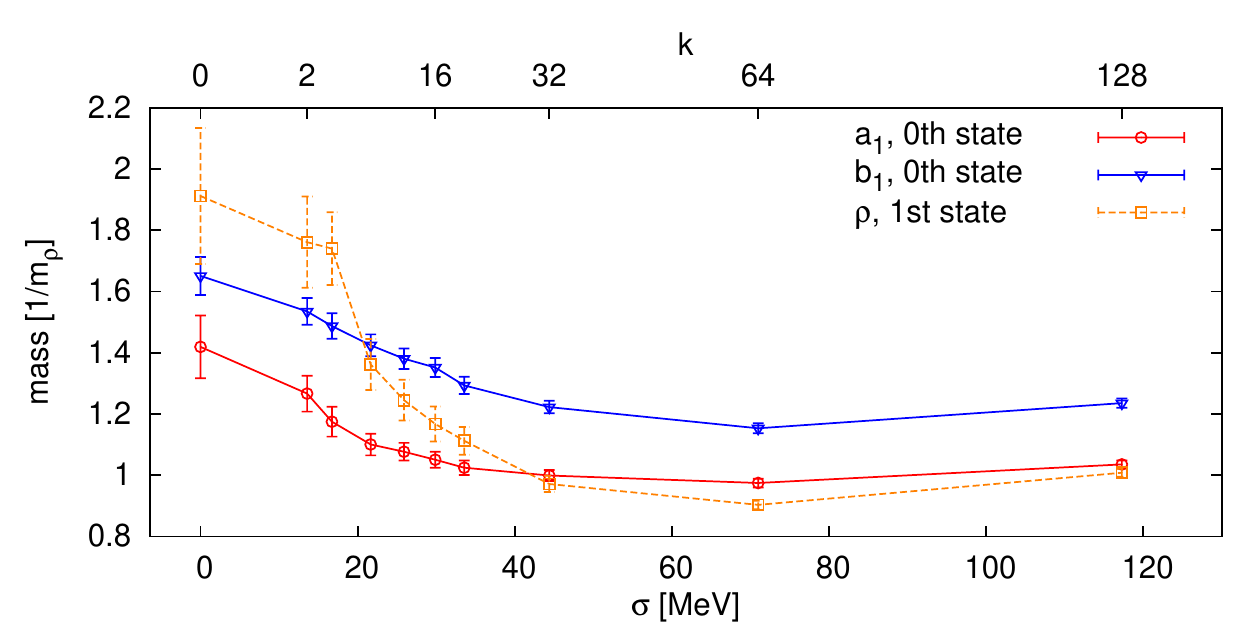}
	\caption{Summary plots: Baryons (l.h.s.) and mesons (r.h.s.) in units of
        the $\rho$-mass at the corresponding truncation level.}
\label{fig:summary_ratios}
\end{figure*}

On top of the chirally symmetric vacuum  the $\rho-a_1$ splitting
vanishes, see, e.g., Figs.  \ref{fig:summary} and
\ref{fig:summary_ratios}., a clear signal of the chiral $SU(2)_L \times
SU(2)_R$ symmetry restoration in the physical states. At the same time 
large $b_1-\rho$ and $b_1-\rho'$ splittings persist. This is a
direct indication that the $U(1)_A$ breaking does not disappear. While
that $U(1)_A$ breaking component that is due to the chiral  condensate
should vanish with the condensate, the $U(1)_A$ breaking via the axial
anomaly still persists. Then it follows that there is no direct
interconnection of the lowest lying modes of the Dirac operator and the
mechanism of the anomalous $U(1)_A$ breaking in QCD. Such a direct
interconnection was suggested in the past through, e.g., the instanton
fluctuations.

After unbreaking of the chiral symmetry the $\rho$ and $\rho'$ mesons
become degenerate. What does this tell us? A degeneracy indicates some
symmetry. The two distinct $\rho$ states, $\rho$ and $\rho'$, lie in
different irreducible parity-chiral representations, $(\otf,\otf)_a$ and
$(0,1)+(1,0)$. In principle, their degeneracy could point out to a
reducible representation of the parity-chiral group that would include
both irreducible representations. Indeed, the product of two fundamental
quark-antiquark chiral representations does contain, in particular,  both
$(\otf,\otf)_a$ and $(0,1)+(1,0)$:
\begin{align}
 &[(0,\otf)+(\otf,0)] \times [(0,\otf)+(\otf,0)]\nonumber\\
  &\;\;= (0,0) + (\otf,\otf)_a
 + (\otf,\otf)_b +  (0,1)+(1,0)\FD
\label{barqq}
\end{align}
Such a multiplet of dimension 16 (including isospin degeneracies) would
consist of  two distinct $\omega$-mesons, $f_1$, $h_1$, two $\rho$-mesons
as well as $b_1$ and $a_1$ mesons and would require a degeneracy of all
of them. Now we do find, however, that the $b_1$ meson is well split from
both $\rho$ and $\rho'$ after the unbreaking of the chiral symmetry. This
rules out that the observed $\rho -\rho'$ degeneracy is related to
restored chiral symmetry. The degenerate $\rho$ and $\rho'$ states 
are different because
their eigenvectors are orthogonal and because they are well split before the
removal of the low modes. This can be clearly seen from Figs. 
\ref{fig:summary} and \ref{fig:summary_ratios}.
This degeneracy indicates some higher symmetry
that includes chiral $SU(2)_L \times SU(2)_R$ as a subgroup. It is a
highly exciting question what this higher symmetry is. It will be seen
from  the following subsection that baryons also point to some higher
symmetry.


\subsection{Baryon chiral multiplets}

If chiral symmetry is restored and baryons are still there they have to
fall into (some of) the possible baryonic parity-chiral multiplets. There
are three different irreducible representations  of $SU(2)_L \times
SU(2)_R \times C_i$ for baryons of any fixed spin:
\begin{equation}
(\otf,0) + (0,\otf)\,,~(\ttf,0) + (0,\ttf)\,,~ (\otf,1) + (1,\otf)\,. 
\label{bm}
\end{equation}
The first representation combines nucleons of positive and negative
parity into a parity doublet. The second representation consists of both
positive and negative parity $\Delta$'s of the same spin. Finally, the
third representation, that is a quartet, includes one nucleon and one
Delta parity doublet with the same spin.

Extraction of the chiral eigenmodes of the Dirac operator leads to a
systematic appearance of the parity doublets, as it is clearly seen from
Figs. \ref{fig:summary} and \ref{fig:summary_ratios}. There are two
degenerate nucleon parity doublets with the same mass. There are also two
distinct $\Delta$ parity doublets, but with different mass. Since our
interpolators have spin $J=\otf$ for nucleons and $J=\ttf$ for Delta's,
we cannot see possible quartets of the  $(\otf,1) + (1,\otf)$ type.

It is very interesting that the two nucleon parity doublets get
degenerate, while the two Delta doublets are well split. The former hints
at a higher symmetry for the $J=I=\otf$ states, while this higher
symmetry is absent for the $J=I=\ttf$ states.

\subsection{On the origin of the hyperfine splitting in QCD}

The $\Delta-N$ splitting is usually attributed to the hyperfine
spin-spin interaction between valence quarks. The realistic candidates
for this interaction are the spin-spin color-magnetic interaction
\cite{De_Rujula:1975ge,Isgur:1978xj} and the flavor-spin interaction
related to the spontaneous chiral symmetry breaking
\cite{Glozman:1995fu}. It is an old debated issue which one is really
responsible for the hyperfine splittings in baryons. Our results suggest
some answer to this question. Once chiral symmetry breaking is removed,
which happens for the ground $N$ and $\Delta$ states after extraction of
the 50--60 lowest eigenmodes, the $\Delta-N$ splitting is reduced
roughly by  the factor 2. With the restored chiral symmetry  the
effective flavor-spin quark-quark interaction is impossible. The
color-magnetic interaction is still there. This result suggests that in
our real world the contribution of both these mechanisms to the
$\Delta-N$ splitting is of equal importance.
 
\section{Conclusions}\label{sec:conclusion}

We have studied what happens with different mesons and baryons upon
modifying the valence quark propagators by removing the lowest lying
eigenmodes of the Dirac operator. These eigenmodes are directly related
to the quark condensate of the vacuum via the Banks--Casher relation.
Consequently, upon removal of the lowest eigenmodes we artificially
restore chiral symmetry, what we call ``unbreaking'' of the chiral
symmetry. We study the evolution of the hadron masses with the number of
extracted lowest eigenmodes in dynamical lattice simulations. There are a few
interesting observations.

First, the quality of the signals from the hadrons after removal of the
chiral eigenmodes of the valence quark propagators become  much better
than with the untruncated propagators. Most probably this is related to
the fact that we artificially remove the pion cloud of the hadrons.

Second, from the spectral patterns both for the ground and excited mesons
and baryons we conclude that confinement is still there while the chiral
symmetry is artificially restored. Restoration of the chiral symmetry is
evidenced by the fact that hadrons, both baryons and mesons, fall into
different parity-chiral multiplets. At the same time there is a clear
evidence that the broken $U(1)_A$ symmetry is not restored. 

Third, some distinct  parity doublets get degenerate upon chiral symmetry
restoration. This indicates that there is some higher symmetry in the
chirally restored regime, that includes the chiral group as a subgroup.

Finally, from the comparison of the hyperfine $\Delta - N$ splitting
before and after unbreaking of the chiral symmetry we conclude that in
our real world both the color-magnetic and the flavor-spin interactions
between valence quarks are of equal importance.

\begin{acknowledgments}
We would like to thank Georg P. Engel and Kim Splittorff for valuable discussions.  The
calculations have been performed on the SGI Altix 4700 of the LRZ Munich
and on clusters at ZID at the University of Graz. Support by DFG SFB-TR55
and by the Austrian Science Fund (FWF) through grants P21970-N16 and DK
W1203-N16 is gratefully acknowledged. M.S. is supported by the Research
Executive  Agency (REA) of the European Union under Grant Agreement 
PITN-GA-2009-238353 (ITN STRONGnet).

\end{acknowledgments}


%


\begin{thebibliography}{55}%
\makeatletter
\providecommand \@ifxundefined [1]{%
 \@ifx{#1\undefined}
}%
\providecommand \@ifnum [1]{%
 \ifnum #1\expandafter \@firstoftwo
 \else \expandafter \@secondoftwo
 \fi
}%
\providecommand \@ifx [1]{%
 \ifx #1\expandafter \@firstoftwo
 \else \expandafter \@secondoftwo
 \fi
}%
\providecommand \natexlab [1]{#1}%
\providecommand \enquote  [1]{``#1''}%
\providecommand \bibnamefont  [1]{#1}%
\providecommand \bibfnamefont [1]{#1}%
\providecommand \citenamefont [1]{#1}%
\providecommand \href@noop [0]{\@secondoftwo}%
\providecommand \href [0]{\begingroup \@sanitize@url \@href}%
\providecommand \@href[1]{\@@startlink{#1}\@@href}%
\providecommand \@@href[1]{\endgroup#1\@@endlink}%
\providecommand \@sanitize@url [0]{\catcode `\\12\catcode `\$12\catcode
  `\&12\catcode `\#12\catcode `\^12\catcode `\_12\catcode `\%12\relax}%
\providecommand \@@startlink[1]{}%
\providecommand \@@endlink[0]{}%
\providecommand \url  [0]{\begingroup\@sanitize@url \@url }%
\providecommand \@url [1]{\endgroup\@href {#1}{\urlprefix }}%
\providecommand \urlprefix  [0]{URL }%
\providecommand \Eprint [0]{\href }%
\providecommand \doibase [0]{http://dx.doi.org/}%
\providecommand \selectlanguage [0]{\@gobble}%
\providecommand \bibinfo  [0]{\@secondoftwo}%
\providecommand \bibfield  [0]{\@secondoftwo}%
\providecommand \translation [1]{[#1]}%
\providecommand \BibitemOpen [0]{}%
\providecommand \bibitemStop [0]{}%
\providecommand \bibitemNoStop [0]{.\EOS\space}%
\providecommand \EOS [0]{\spacefactor3000\relax}%
\providecommand \BibitemShut  [1]{\csname bibitem#1\endcsname}%
\let\auto@bib@innerbib\@empty
\bibitem [{\citenamefont {Glozman}(2000)}]{Glozman:1999tk}%
  \BibitemOpen
  \bibfield  {author} {\bibinfo {author} {\bibfnamefont {L.~Y.}\ \bibnamefont
  {Glozman}},\ }\href {\doibase 10.1016/S0370-2693(00)00096-4} {\bibfield
  {journal} {\bibinfo  {journal} {Phys. Lett. B}\ }\textbf {\bibinfo {volume}
  {475}},\ \bibinfo {pages} {329} (\bibinfo {year} {2000})},\ \Eprint
  {http://arxiv.org/abs/hep-ph/9908207} {arXiv:hep-ph/9908207} \BibitemShut
  {NoStop}%
\bibitem [{\citenamefont {Cohen}\ and\ \citenamefont
  {Glozman}(2001)}]{Cohen:2001gb}%
  \BibitemOpen
  \bibfield  {author} {\bibinfo {author} {\bibfnamefont {T.~D.}\ \bibnamefont
  {Cohen}}\ and\ \bibinfo {author} {\bibfnamefont {L.~Y.}\ \bibnamefont
  {Glozman}},\ }\href {\doibase 10.1103/PhysRevD.65.016006} {\bibfield
  {journal} {\bibinfo  {journal} {Phys. Rev. D}\ }\textbf {\bibinfo {volume}
  {65}},\ \bibinfo {pages} {016006} (\bibinfo {year} {2001})},\ \Eprint
  {http://arxiv.org/abs/hep-ph/0102206} {arXiv:hep-ph/0102206} \BibitemShut
  {NoStop}%
\bibitem [{\citenamefont {Cohen}\ and\ \citenamefont
  {Glozman}(2002)}]{Cohen:2002st}%
  \BibitemOpen
  \bibfield  {author} {\bibinfo {author} {\bibfnamefont {T.~D.}\ \bibnamefont
  {Cohen}}\ and\ \bibinfo {author} {\bibfnamefont {L.~Y.}\ \bibnamefont
  {Glozman}},\ }\href {\doibase 10.1142/S0217751X02009679} {\bibfield
  {journal} {\bibinfo  {journal} {Int. J. Mod. Phys.}\ }\textbf {\bibinfo
  {volume} {A17}},\ \bibinfo {pages} {1327} (\bibinfo {year} {2002})},\ \Eprint
  {http://arxiv.org/abs/hep-ph/0201242} {arXiv:hep-ph/0201242} \BibitemShut
  {NoStop}%
\bibitem [{\citenamefont {Glozman}(2002)}]{Glozman:2002cp}%
  \BibitemOpen
  \bibfield  {author} {\bibinfo {author} {\bibfnamefont {L.~Y.}\ \bibnamefont
  {Glozman}},\ }\href {\doibase 10.1016/S0370-2693(02)02096-8} {\bibfield
  {journal} {\bibinfo  {journal} {Phys. Lett. B}\ }\textbf {\bibinfo {volume}
  {539}},\ \bibinfo {pages} {257} (\bibinfo {year} {2002})},\ \Eprint
  {http://arxiv.org/abs/hep-ph/0205072} {arXiv:hep-ph/0205072 [hep-ph]}
  \BibitemShut {NoStop}%
\bibitem [{\citenamefont {Glozman}(2004)}]{Glozman:2003bt}%
  \BibitemOpen
  \bibfield  {author} {\bibinfo {author} {\bibfnamefont {L.~Y.}\ \bibnamefont
  {Glozman}},\ }\href {\doibase 10.1016/j.physletb.2004.02.066} {\bibfield
  {journal} {\bibinfo  {journal} {Phys. Lett. B}\ }\textbf {\bibinfo {volume}
  {587}},\ \bibinfo {pages} {69} (\bibinfo {year} {2004})},\ \Eprint
  {http://arxiv.org/abs/hep-ph/0312354} {arXiv:hep-ph/0312354} \BibitemShut
  {NoStop}%
\bibitem [{\citenamefont {Glozman}(2007)}]{Glozman:2007ek}%
  \BibitemOpen
  \bibfield  {author} {\bibinfo {author} {\bibfnamefont {L.~Y.}\ \bibnamefont
  {Glozman}},\ }\href@noop {} {\bibfield  {journal} {\bibinfo  {journal} {Phys.
  Rep.}\ }\textbf {\bibinfo {volume} {444}},\ \bibinfo {pages} {1} (\bibinfo
  {year} {2007})},\ \Eprint {http://arxiv.org/abs/hep-ph/0701081}
  {hep-ph/0701081} \BibitemShut {NoStop}%
\bibitem [{\citenamefont {Jaffe}\ \emph
  {et~al.}(2006{\natexlab{a}})\citenamefont {Jaffe}, \citenamefont {Pirjol},\
  and\ \citenamefont {Scardicchio}}]{Jaffe:2006aq}%
  \BibitemOpen
  \bibfield  {author} {\bibinfo {author} {\bibfnamefont {R.~L.}\ \bibnamefont
  {Jaffe}}, \bibinfo {author} {\bibfnamefont {D.}~\bibnamefont {Pirjol}}, \
  and\ \bibinfo {author} {\bibfnamefont {A.}~\bibnamefont {Scardicchio}},\
  }\href {\doibase 10.1103/PhysRevD.74.057901} {\bibfield  {journal} {\bibinfo
  {journal} {Phys. Rev. D}\ }\textbf {\bibinfo {volume} {74}},\ \bibinfo
  {pages} {057901} (\bibinfo {year} {2006}{\natexlab{a}})}\BibitemShut
  {NoStop}%
\bibitem [{\citenamefont {Jaffe}\ \emph
  {et~al.}(2006{\natexlab{b}})\citenamefont {Jaffe}, \citenamefont {Pirjol},\
  and\ \citenamefont {Scardicchio}}]{Jaffe:2005sq}%
  \BibitemOpen
  \bibfield  {author} {\bibinfo {author} {\bibfnamefont {R.~L.}\ \bibnamefont
  {Jaffe}}, \bibinfo {author} {\bibfnamefont {D.}~\bibnamefont {Pirjol}}, \
  and\ \bibinfo {author} {\bibfnamefont {A.}~\bibnamefont {Scardicchio}},\
  }\href {\doibase 10.1103/PhysRevLett.96.121601} {\bibfield  {journal}
  {\bibinfo  {journal} {Phys. Rev. Lett.}\ }\textbf {\bibinfo {volume} {96}},\
  \bibinfo {pages} {121601} (\bibinfo {year} {2006}{\natexlab{b}})},\ \Eprint
  {http://arxiv.org/abs/hep-ph/0511081} {arXiv:hep-ph/0511081} \BibitemShut
  {NoStop}%
\bibitem [{\citenamefont {Jaffe}\ \emph
  {et~al.}(2006{\natexlab{c}})\citenamefont {Jaffe}, \citenamefont {Pirjol},\
  and\ \citenamefont {Scardicchio}}]{Jaffe:2006jy}%
  \BibitemOpen
  \bibfield  {author} {\bibinfo {author} {\bibfnamefont {R.~L.}\ \bibnamefont
  {Jaffe}}, \bibinfo {author} {\bibfnamefont {D.}~\bibnamefont {Pirjol}}, \
  and\ \bibinfo {author} {\bibfnamefont {A.}~\bibnamefont {Scardicchio}},\
  }\href {\doibase 10.1016/j.physrep.2006.09.004} {\bibfield  {journal}
  {\bibinfo  {journal} {Phys. Rept.}\ }\textbf {\bibinfo {volume} {435}},\
  \bibinfo {pages} {157} (\bibinfo {year} {2006}{\natexlab{c}})},\ \Eprint
  {http://arxiv.org/abs/hep-ph/0602010} {arXiv:hep-ph/0602010} \BibitemShut
  {NoStop}%
\bibitem [{\citenamefont {Shifman}\ and\ \citenamefont
  {Vainshtein}(2008)}]{Shifman:2007xn}%
  \BibitemOpen
  \bibfield  {author} {\bibinfo {author} {\bibfnamefont {M.}~\bibnamefont
  {Shifman}}\ and\ \bibinfo {author} {\bibfnamefont {A.}~\bibnamefont
  {Vainshtein}},\ }\href {\doibase 10.1103/PhysRevD.77.034002} {\bibfield
  {journal} {\bibinfo  {journal} {Phys. Rev. D}\ }\textbf {\bibinfo {volume}
  {77}},\ \bibinfo {pages} {034002} (\bibinfo {year} {2008})},\ \Eprint
  {http://arxiv.org/abs/0710.0863} {arXiv:0710.0863 [hep-ph]} \BibitemShut
  {NoStop}%
\bibitem [{\citenamefont {Anisovich}\ \emph {et~al.}(2012)\citenamefont
  {Anisovich}, \citenamefont {Klempt}, \citenamefont {Nikonov}, \citenamefont
  {Sarantsev}, \citenamefont {Schmieden} \emph {et~al.}}]{Anisovich:2011sv}%
  \BibitemOpen
  \bibfield  {author} {\bibinfo {author} {\bibfnamefont {A.}~\bibnamefont
  {Anisovich}}, \bibinfo {author} {\bibfnamefont {E.}~\bibnamefont {Klempt}},
  \bibinfo {author} {\bibfnamefont {V.}~\bibnamefont {Nikonov}}, \bibinfo
  {author} {\bibfnamefont {A.}~\bibnamefont {Sarantsev}}, \bibinfo {author}
  {\bibfnamefont {H.}~\bibnamefont {Schmieden}},  \emph {et~al.},\ }\href@noop
  {} {\bibfield  {journal} {\bibinfo  {journal} {Phys. Lett. B}\ }\textbf
  {\bibinfo {volume} {711}},\ \bibinfo {pages} {162} (\bibinfo {year}
  {2012})},\ \Eprint {http://arxiv.org/abs/1111.6151} {arXiv:1111.6151
  [nucl-ex]} \BibitemShut {NoStop}%
\bibitem [{\citenamefont {DeGrand}(2004)}]{DeGrand:2003sf}%
  \BibitemOpen
  \bibfield  {author} {\bibinfo {author} {\bibfnamefont {T.~A.}\ \bibnamefont
  {DeGrand}},\ }\href {\doibase 10.1103/PhysRevD.69.074024} {\bibfield
  {journal} {\bibinfo  {journal} {Phys. Rev. D}\ }\textbf {\bibinfo {volume}
  {69}},\ \bibinfo {pages} {074024} (\bibinfo {year} {2004})},\ \Eprint
  {http://arxiv.org/abs/hep-ph/0310303} {arXiv:hep-ph/0310303 [hep-ph]}
  \BibitemShut {NoStop}%
\bibitem [{\citenamefont {Cohen}(2006)}]{Cohen:2006bq}%
  \BibitemOpen
  \bibfield  {author} {\bibinfo {author} {\bibfnamefont {T.~D.}\ \bibnamefont
  {Cohen}},\ }\href {\doibase 10.1016/j.nuclphysa.2006.06.015} {\bibfield
  {journal} {\bibinfo  {journal} {Nucl.Phys.}\ }\textbf {\bibinfo {volume}
  {A775}},\ \bibinfo {pages} {89} (\bibinfo {year} {2006})},\ \Eprint
  {http://arxiv.org/abs/hep-ph/0605206} {arXiv:hep-ph/0605206 [hep-ph]}
  \BibitemShut {NoStop}%
\bibitem [{\citenamefont {Sch{{\"a}}fer}\ and\ \citenamefont
  {Shuryak}(1998)}]{Schafer:1996wv}%
  \BibitemOpen
  \bibfield  {author} {\bibinfo {author} {\bibfnamefont {T.}~\bibnamefont
  {Sch{{\"a}}fer}}\ and\ \bibinfo {author} {\bibfnamefont {E.~V.}\ \bibnamefont
  {Shuryak}},\ }\href@noop {} {\bibfield  {journal} {\bibinfo  {journal} {Rev.
  Mod. Phys.}\ }\textbf {\bibinfo {volume} {70}},\ \bibinfo {pages} {323}
  (\bibinfo {year} {1998})},\ \Eprint {http://arxiv.org/abs/hep-ph/9610451}
  {hep-ph/9610451} \BibitemShut {NoStop}%
\bibitem [{\citenamefont {DeGrand}\ and\ \citenamefont
  {Hasenfratz}(2001)}]{DeGrand:2000gq}%
  \BibitemOpen
  \bibfield  {author} {\bibinfo {author} {\bibfnamefont {T.~A.}\ \bibnamefont
  {DeGrand}}\ and\ \bibinfo {author} {\bibfnamefont {A.}~\bibnamefont
  {Hasenfratz}},\ }\href {\doibase 10.1103/PhysRevD.64.034512} {\bibfield
  {journal} {\bibinfo  {journal} {Phys. Rev. D}\ }\textbf {\bibinfo {volume}
  {64}},\ \bibinfo {pages} {034512} (\bibinfo {year} {2001})},\ \Eprint
  {http://arxiv.org/abs/hep-lat/0012021} {arXiv:hep-lat/0012021 [hep-lat]}
  \BibitemShut {NoStop}%
\bibitem [{\citenamefont {Neff}\ \emph {et~al.}(2001)\citenamefont {Neff},
  \citenamefont {Eicker}, \citenamefont {Lippert}, \citenamefont {Negele},\
  and\ \citenamefont {Schilling}}]{Neff:2001zr}%
  \BibitemOpen
  \bibfield  {author} {\bibinfo {author} {\bibfnamefont {H.}~\bibnamefont
  {Neff}}, \bibinfo {author} {\bibfnamefont {N.}~\bibnamefont {Eicker}},
  \bibinfo {author} {\bibfnamefont {T.}~\bibnamefont {Lippert}}, \bibinfo
  {author} {\bibfnamefont {J.~W.}\ \bibnamefont {Negele}}, \ and\ \bibinfo
  {author} {\bibfnamefont {K.}~\bibnamefont {Schilling}},\ }\href {\doibase
  10.1103/PhysRevD.64.114509} {\bibfield  {journal} {\bibinfo  {journal}
  {Phys.Rev.}\ }\textbf {\bibinfo {volume} {D64}},\ \bibinfo {pages} {114509}
  (\bibinfo {year} {2001})},\ \Eprint {http://arxiv.org/abs/hep-lat/0106016}
  {arXiv:hep-lat/0106016 [hep-lat]} \BibitemShut {NoStop}%
\bibitem [{\citenamefont {DeGrand}(2001)}]{DeGrand:2001tm}%
  \BibitemOpen
  \bibfield  {author} {\bibinfo {author} {\bibfnamefont {T.}~\bibnamefont
  {DeGrand}},\ }\href@noop {} {\bibfield  {journal} {\bibinfo  {journal} {Phys.
  Rev. D}\ }\textbf {\bibinfo {volume} {64}},\ \bibinfo {pages} {094508}
  (\bibinfo {year} {2001})},\ \Eprint {http://arxiv.org/abs/hep-lat/0106001}
  {hep-lat/0106001 [hep-lat]} \BibitemShut {NoStop}%
\bibitem [{\citenamefont {DeGrand}\ and\ \citenamefont
  {Schaefer}(2004)}]{DeGrand:2004qw}%
  \BibitemOpen
  \bibfield  {author} {\bibinfo {author} {\bibfnamefont {T.}~\bibnamefont
  {DeGrand}}\ and\ \bibinfo {author} {\bibfnamefont {S.}~\bibnamefont
  {Schaefer}},\ }\href@noop {} {\bibfield  {journal} {\bibinfo  {journal}
  {Comput. Phys. Commun.}\ }\textbf {\bibinfo {volume} {159}},\ \bibinfo
  {pages} {185} (\bibinfo {year} {2004})},\ \Eprint
  {http://arxiv.org/abs/hep-lat/0401011} {hep-lat/0401011} \BibitemShut
  {NoStop}%
\bibitem [{\citenamefont {Neuberger}(1998{\natexlab{a}})}]{Neuberger:1997fp}%
  \BibitemOpen
  \bibfield  {author} {\bibinfo {author} {\bibfnamefont {H.}~\bibnamefont
  {Neuberger}},\ }\href {\doibase 10.1016/S0370-2693(97)01368-3} {\bibfield
  {journal} {\bibinfo  {journal} {Phys. Lett. B}\ }\textbf {\bibinfo {volume}
  {417}},\ \bibinfo {pages} {141} (\bibinfo {year} {1998}{\natexlab{a}})},\
  \Eprint {http://arxiv.org/abs/hep-lat/9707022} {arXiv:hep-lat/9707022}
  \BibitemShut {NoStop}%
\bibitem [{\citenamefont {Neuberger}(1998{\natexlab{b}})}]{Neuberger:1998wv}%
  \BibitemOpen
  \bibfield  {author} {\bibinfo {author} {\bibfnamefont {H.}~\bibnamefont
  {Neuberger}},\ }\href {\doibase 10.1016/S0370-2693(98)00355-4} {\bibfield
  {journal} {\bibinfo  {journal} {Phys. Lett. B}\ }\textbf {\bibinfo {volume}
  {427}},\ \bibinfo {pages} {353} (\bibinfo {year} {1998}{\natexlab{b}})},\
  \Eprint {http://arxiv.org/abs/hep-lat/9801031} {arXiv:hep-lat/9801031}
  \BibitemShut {NoStop}%
\bibitem [{\citenamefont {DeGrand}\ and\ \citenamefont
  {Schaefer}(2005)}]{DeGrand:2004wh}%
  \BibitemOpen
  \bibfield  {author} {\bibinfo {author} {\bibfnamefont {T.}~\bibnamefont
  {DeGrand}}\ and\ \bibinfo {author} {\bibfnamefont {S.}~\bibnamefont
  {Schaefer}},\ }\href@noop {} {\bibfield  {journal} {\bibinfo  {journal}
  {Nucl. Phys. (Proc. Suppl.)}\ }\textbf {\bibinfo {volume} {140}},\ \bibinfo
  {pages} {296} (\bibinfo {year} {2005})},\ \Eprint
  {http://arxiv.org/abs/hep-lat/0409056} {hep-lat/0409056} \BibitemShut
  {NoStop}%
\bibitem [{\citenamefont {Giusti}\ \emph {et~al.}(2004)\citenamefont {Giusti},
  \citenamefont {Hern\'andez}, \citenamefont {Laine}, \citenamefont {Weisz},\
  and\ \citenamefont {Wittig}}]{Giusti:2004yp}%
  \BibitemOpen
  \bibfield  {author} {\bibinfo {author} {\bibfnamefont {L.}~\bibnamefont
  {Giusti}}, \bibinfo {author} {\bibfnamefont {P.}~\bibnamefont {Hern\'andez}},
  \bibinfo {author} {\bibfnamefont {M.}~\bibnamefont {Laine}}, \bibinfo
  {author} {\bibfnamefont {P.}~\bibnamefont {Weisz}}, \ and\ \bibinfo {author}
  {\bibfnamefont {H.}~\bibnamefont {Wittig}},\ }\href@noop {} {\bibfield
  {journal} {\bibinfo  {journal} {JHEP}\ }\textbf {\bibinfo {volume} {04}},\
  \bibinfo {pages} {013} (\bibinfo {year} {2004})},\ \Eprint
  {http://arxiv.org/abs/hep-lat/0402002} {hep-lat/0402002} \BibitemShut
  {NoStop}%
\bibitem [{\citenamefont {Bali}\ \emph
  {et~al.}(2010{\natexlab{a}})\citenamefont {Bali}, \citenamefont {Collins},\
  and\ \citenamefont {Schaefer}}]{Bali:2009hu}%
  \BibitemOpen
  \bibfield  {author} {\bibinfo {author} {\bibfnamefont {G.~S.}\ \bibnamefont
  {Bali}}, \bibinfo {author} {\bibfnamefont {S.}~\bibnamefont {Collins}}, \
  and\ \bibinfo {author} {\bibfnamefont {A.}~\bibnamefont {Schaefer}},\
  }\href@noop {} {\bibfield  {journal} {\bibinfo  {journal} {Comput. Phys.
  Commun.}\ }\textbf {\bibinfo {volume} {181}},\ \bibinfo {pages} {1570}
  (\bibinfo {year} {2010}{\natexlab{a}})},\ \Eprint
  {http://arxiv.org/abs/0910.3970} {0910.3970} \BibitemShut {NoStop}%
\bibitem [{\citenamefont {Bali}\ \emph
  {et~al.}(2010{\natexlab{b}})\citenamefont {Bali}, \citenamefont
  {Castagnini},\ and\ \citenamefont {Collins}}]{Bali:2010se}%
  \BibitemOpen
  \bibfield  {author} {\bibinfo {author} {\bibfnamefont {G.}~\bibnamefont
  {Bali}}, \bibinfo {author} {\bibfnamefont {L.}~\bibnamefont {Castagnini}}, \
  and\ \bibinfo {author} {\bibfnamefont {S.}~\bibnamefont {Collins}},\
  }\href@noop {} {\bibfield  {journal} {\bibinfo  {journal} {PoS}\ }\textbf
  {\bibinfo {volume} {LATTICE2010}},\ \bibinfo {pages} {096} (\bibinfo {year}
  {2010}{\natexlab{b}})},\ \Eprint {http://arxiv.org/abs/1011.1353}
  {arXiv:1011.1353 [hep-lat]} \BibitemShut {NoStop}%
\bibitem [{\citenamefont {Lang}\ and\ \citenamefont
  {Schr{\"o}ck}(2011{\natexlab{a}})}]{Lang:2011qy}%
  \BibitemOpen
  \bibfield  {author} {\bibinfo {author} {\bibfnamefont {C.~B.}\ \bibnamefont
  {Lang}}\ and\ \bibinfo {author} {\bibfnamefont {M.}~\bibnamefont
  {Schr{\"o}ck}},\ }\href@noop {} {\bibfield  {journal} {\bibinfo  {journal}
  {Phys. Rev. D}\ }\textbf {\bibinfo {volume} {84}},\ \bibinfo {pages} {087704}
  (\bibinfo {year} {2011}{\natexlab{a}})},\ \Eprint
  {http://arxiv.org/abs/1107.5195} {1107.5195 [hep-lat]} \BibitemShut {NoStop}%
\bibitem [{\citenamefont {Lang}\ and\ \citenamefont
  {Schr{\"o}ck}(2011{\natexlab{b}})}]{Lang:2011ai}%
  \BibitemOpen
  \bibfield  {author} {\bibinfo {author} {\bibfnamefont {C.~B.}\ \bibnamefont
  {Lang}}\ and\ \bibinfo {author} {\bibfnamefont {M.}~\bibnamefont
  {Schr{\"o}ck}},\ }\href@noop {} {\bibfield  {journal} {\bibinfo  {journal}
  {PoS}\ }\textbf {\bibinfo {volume} {LATTICE2011}},\ \bibinfo {pages} {111}
  (\bibinfo {year} {2011}{\natexlab{b}})},\ \Eprint
  {http://arxiv.org/abs/1110.6149} {arXiv:1110.6149 [hep-lat]} \BibitemShut
  {NoStop}%
\bibitem [{\citenamefont {Schr{\"o}ck}(2012)}]{Schrock:2011hq}%
  \BibitemOpen
  \bibfield  {author} {\bibinfo {author} {\bibfnamefont {M.}~\bibnamefont
  {Schr{\"o}ck}},\ }\href@noop {} {\bibfield  {journal} {\bibinfo  {journal}
  {Phys. Lett. B}\ }\textbf {\bibinfo {volume} {711}},\ \bibinfo {pages} {217}
  (\bibinfo {year} {2012})},\ \Eprint {http://arxiv.org/abs/1112.5107}
  {arXiv:1112.5107 [hep-lat]} \BibitemShut {NoStop}%
\bibitem [{\citenamefont {Suganuma}\ \emph {et~al.}(2011)\citenamefont
  {Suganuma}, \citenamefont {Gongyo}, \citenamefont {Iritani},\ and\
  \citenamefont {Yamamoto}}]{Suganuma:2011kn}%
  \BibitemOpen
  \bibfield  {author} {\bibinfo {author} {\bibfnamefont {H.}~\bibnamefont
  {Suganuma}}, \bibinfo {author} {\bibfnamefont {S.}~\bibnamefont {Gongyo}},
  \bibinfo {author} {\bibfnamefont {T.}~\bibnamefont {Iritani}}, \ and\
  \bibinfo {author} {\bibfnamefont {A.}~\bibnamefont {Yamamoto}},\ }\href@noop
  {} {\bibfield  {journal} {\bibinfo  {journal} {PoS}\ }\textbf {\bibinfo
  {volume} {QCD-TNT-II}},\ \bibinfo {pages} {044} (\bibinfo {year} {2011})},\
  \Eprint {http://arxiv.org/abs/1112.1962} {arXiv:1112.1962 [hep-lat]}
  \BibitemShut {NoStop}%
\bibitem [{\citenamefont {Gongyo}\ \emph {et~al.}(2012)\citenamefont {Gongyo},
  \citenamefont {Iritani},\ and\ \citenamefont {Suganuma}}]{Gongyo:2012vx}%
  \BibitemOpen
  \bibfield  {author} {\bibinfo {author} {\bibfnamefont {S.}~\bibnamefont
  {Gongyo}}, \bibinfo {author} {\bibfnamefont {T.}~\bibnamefont {Iritani}}, \
  and\ \bibinfo {author} {\bibfnamefont {H.}~\bibnamefont {Suganuma}},\ }\href
  {http://arxiv.org/abs/1202.4130v1} {\enquote {\bibinfo {title}
  {Gauge-invariant formalism with Dirac-mode expansion for confinement and
  chiral symmetry breaking},}\ } (\bibinfo {year} {2012}),\ \Eprint
  {http://arxiv.org/abs/1202.4130v1} {1202.4130v1} \BibitemShut {NoStop}%
\bibitem [{\citenamefont {Brown}\ and\ \citenamefont
  {Rho}(1991)}]{Brown:1991kk}%
  \BibitemOpen
  \bibfield  {author} {\bibinfo {author} {\bibfnamefont {G.~E.}\ \bibnamefont
  {Brown}}\ and\ \bibinfo {author} {\bibfnamefont {M.}~\bibnamefont {Rho}},\
  }\href {\doibase 10.1103/PhysRevLett.66.2720} {\bibfield  {journal} {\bibinfo
   {journal} {Phys. Rev. Lett.}\ }\textbf {\bibinfo {volume} {66}},\ \bibinfo
  {pages} {2720} (\bibinfo {year} {1991})}\BibitemShut {NoStop}%
\bibitem [{\citenamefont {Glozman}\ and\ \citenamefont
  {Wagenbrunn}(2008)}]{Glozman:2007tv}%
  \BibitemOpen
  \bibfield  {author} {\bibinfo {author} {\bibfnamefont {L.~Y.}\ \bibnamefont
  {Glozman}}\ and\ \bibinfo {author} {\bibfnamefont {R.~F.}\ \bibnamefont
  {Wagenbrunn}},\ }\href {\doibase 10.1103/PhysRevD.77.054027} {\bibfield
  {journal} {\bibinfo  {journal} {Phys. Rev. D}\ }\textbf {\bibinfo {volume}
  {77}},\ \bibinfo {pages} {054027} (\bibinfo {year} {2008})},\ \Eprint
  {http://arxiv.org/abs/0709.3080} {arXiv:0709.3080 [hep-ph]} \BibitemShut
  {NoStop}%
\bibitem [{\citenamefont {Glozman}(2009{\natexlab{a}})}]{Glozman:2008fk}%
  \BibitemOpen
  \bibfield  {author} {\bibinfo {author} {\bibfnamefont {L.~Y.}\ \bibnamefont
  {Glozman}},\ }\href {\doibase 10.1103/PhysRevD.79.037504} {\bibfield
  {journal} {\bibinfo  {journal} {Phys. Rev. D}\ }\textbf {\bibinfo {volume}
  {79}},\ \bibinfo {pages} {037504} (\bibinfo {year} {2009}{\natexlab{a}})},\
  \Eprint {http://arxiv.org/abs/0812.1101} {arXiv:0812.1101 [hep-ph]}
  \BibitemShut {NoStop}%
\bibitem [{\citenamefont {Glozman}(2009{\natexlab{b}})}]{Glozman:2009sa}%
  \BibitemOpen
  \bibfield  {author} {\bibinfo {author} {\bibfnamefont {L.~Y.}\ \bibnamefont
  {Glozman}},\ }\href {\doibase 10.1103/PhysRevD.80.037701} {\bibfield
  {journal} {\bibinfo  {journal} {Phys. Rev. D}\ }\textbf {\bibinfo {volume}
  {80}},\ \bibinfo {pages} {037701} (\bibinfo {year} {2009}{\natexlab{b}})},\
  \Eprint {http://arxiv.org/abs/0907.1473} {arXiv:0907.1473 [hep-ph]}
  \BibitemShut {NoStop}%
\bibitem [{\citenamefont {Glozman}\ \emph {et~al.}(2011)\citenamefont
  {Glozman}, \citenamefont {Sazonov},\ and\ \citenamefont
  {Wagenbrunn}}]{Glozman:2011eu}%
  \BibitemOpen
  \bibfield  {author} {\bibinfo {author} {\bibfnamefont {L.~Y.}\ \bibnamefont
  {Glozman}}, \bibinfo {author} {\bibfnamefont {V.~K.}\ \bibnamefont
  {Sazonov}}, \ and\ \bibinfo {author} {\bibfnamefont {R.~F.}\ \bibnamefont
  {Wagenbrunn}},\ }\href {\doibase 10.1103/PhysRevD.84.095009} {\bibfield
  {journal} {\bibinfo  {journal} {Phys. Rev.}\ }\textbf {\bibinfo {volume}
  {84}},\ \bibinfo {pages} {095009} (\bibinfo {year} {2011})},\ \Eprint
  {http://arxiv.org/abs/1108.1681} {arXiv:1108.1681 [hep-ph]}\BibitemShut{NoStop}%
\bibitem [{\citenamefont {Banks}\ and\ \citenamefont
  {Casher}(1980)}]{Banks:1979yr}%
  \BibitemOpen
  \bibfield  {author} {\bibinfo {author} {\bibfnamefont {T.}~\bibnamefont
  {Banks}}\ and\ \bibinfo {author} {\bibfnamefont {A.}~\bibnamefont {Casher}},\
  }\href {\doibase 10.1016/0550-3213(80)90255-2} {\bibfield  {journal}
  {\bibinfo  {journal} {Nucl. Phys. B}\ }\textbf {\bibinfo {volume} {169}},\
  \bibinfo {pages} {103} (\bibinfo {year} {1980})}\BibitemShut {NoStop}%
\bibitem [{\citenamefont {Osborn}\ and\ \citenamefont
  {Verbaarschot}(1998)}]{Osborn:1998nm}%
  \BibitemOpen
  \bibfield  {author} {\bibinfo {author} {\bibfnamefont {J.~C.}\ \bibnamefont
  {Osborn}}\ and\ \bibinfo {author} {\bibfnamefont {J.~J.~M.}\ \bibnamefont
  {Verbaarschot}},\ }\href@noop {} {\bibfield  {journal} {\bibinfo  {journal}
  {Phys. Rev. Lett.}\ }\textbf {\bibinfo {volume} {81}},\ \bibinfo {pages}
  {268} (\bibinfo {year} {1998})},\ \Eprint
  {http://arxiv.org/abs/hep-ph/9807490} {hep-ph/9807490} \BibitemShut {NoStop}%
\bibitem [{\citenamefont {L{\"u}scher}(2007)}]{Luscher:2007se}%
  \BibitemOpen
  \bibfield  {author} {\bibinfo {author} {\bibfnamefont {M.}~\bibnamefont
  {L{\"u}scher}},\ }\href {\doibase 10.1088/1126-6708/2007/07/081} {\bibfield
  {journal} {\bibinfo  {journal} {JHEP}\ }\textbf {\bibinfo {volume} {07}},\
  \bibinfo {pages} {081} (\bibinfo {year} {2007})},\ \Eprint
  {http://arxiv.org/abs/0706.2298} {arXiv:0706.2298 [hep-lat]} \BibitemShut
  {NoStop}%
\bibitem [{\citenamefont {Giusti}\ and\ \citenamefont
  {L{\"u}scher}(2009)}]{Giusti:2008vb}%
  \BibitemOpen
  \bibfield  {author} {\bibinfo {author} {\bibfnamefont {L.}~\bibnamefont
  {Giusti}}\ and\ \bibinfo {author} {\bibfnamefont {M.}~\bibnamefont
  {L{\"u}scher}},\ }\href {\doibase 10.1088/1126-6708/2009/03/013} {\bibfield
  {journal} {\bibinfo  {journal} {JHEP}\ }\textbf {\bibinfo {volume} {0903}},\
  \bibinfo {pages} {013} (\bibinfo {year} {2009})},\ \Eprint
  {http://arxiv.org/abs/0812.3638} {arXiv:0812.3638 [hep-lat]} \BibitemShut
  {NoStop}%
\bibitem [{\citenamefont {Necco}\ and\ \citenamefont
  {Shindler}(2011)}]{Necco:2011vx}%
  \BibitemOpen
  \bibfield  {author} {\bibinfo {author} {\bibfnamefont {S.}~\bibnamefont
  {Necco}}\ and\ \bibinfo {author} {\bibfnamefont {A.}~\bibnamefont
  {Shindler}},\ }\href {\doibase 10.1007/JHEP04(2011)031} {\bibfield  {journal}
  {\bibinfo  {journal} {JHEP}\ }\textbf {\bibinfo {volume} {04}},\ \bibinfo
  {pages} {031} (\bibinfo {year} {2011})},\ \Eprint
  {http://arxiv.org/abs/1101.1778} {arXiv:1101.1778 [hep-lat]} \BibitemShut
  {NoStop}%
\bibitem [{\citenamefont {Splittorff}\ and\ \citenamefont
  {Verbaarschot}(2011)}]{Splittorff:2011bj}%
  \BibitemOpen
  \bibfield  {author} {\bibinfo {author} {\bibfnamefont {K.}~\bibnamefont
  {Splittorff}}\ and\ \bibinfo {author} {\bibfnamefont {J.~J.~M.}\ \bibnamefont
  {Verbaarschot}},\ }\href@noop {} {\bibfield  {journal} {\bibinfo  {journal}
  {Phys. Rev. D},\ }\textbf {\bibinfo {volume} {84}},\ \bibinfo {pages}
  {065031} (\bibinfo {year} {2011})},\ \Eprint {http://arxiv.org/abs/1105.6229}
  {arXiv:1105.6229 [hep-lat]} \BibitemShut {NoStop}%
\bibitem [{\citenamefont {Gattringer}(2001)}]{Gattringer:2000js}%
  \BibitemOpen
  \bibfield  {author} {\bibinfo {author} {\bibfnamefont {C.}~\bibnamefont
  {Gattringer}},\ }\href {\doibase 10.1103/PhysRevD.63.114501} {\bibfield
  {journal} {\bibinfo  {journal} {Phys. Rev. D}\ }\textbf {\bibinfo {volume}
  {63}},\ \bibinfo {pages} {114501} (\bibinfo {year} {2001})},\ \Eprint
  {http://arxiv.org/abs/hep-lat/0003005} {arXiv:hep-lat/0003005} \BibitemShut
  {NoStop}%
\bibitem [{\citenamefont {Gattringer}\ \emph {et~al.}(2001)\citenamefont
  {Gattringer}, \citenamefont {Hip},\ and\ \citenamefont
  {Lang}}]{Gattringer:2000qu}%
  \BibitemOpen
  \bibfield  {author} {\bibinfo {author} {\bibfnamefont {C.}~\bibnamefont
  {Gattringer}}, \bibinfo {author} {\bibfnamefont {I.}~\bibnamefont {Hip}}, \
  and\ \bibinfo {author} {\bibfnamefont {C.~B.}\ \bibnamefont {Lang}},\ }\href
  {\doibase 10.1016/S0550-3213(00)00717-3} {\bibfield  {journal} {\bibinfo
  {journal} {Nucl. Phys. B}\ }\textbf {\bibinfo {volume} {597}},\ \bibinfo
  {pages} {451} (\bibinfo {year} {2001})},\ \Eprint
  {http://arxiv.org/abs/hep-lat/0007042} {arXiv:hep-lat/0007042} \BibitemShut
  {NoStop}%
\bibitem [{\citenamefont {Gattringer}\ \emph {et~al.}(2009)\citenamefont
  {Gattringer}, \citenamefont {Hagen}, \citenamefont {Lang}, \citenamefont
  {Limmer}, \citenamefont {Mohler},\ and\ \citenamefont
  {Sch{{\"a}}fer}}]{Gattringer:2008vj}%
  \BibitemOpen
  \bibfield  {author} {\bibinfo {author} {\bibfnamefont {C.}~\bibnamefont
  {Gattringer}}, \bibinfo {author} {\bibfnamefont {C.}~\bibnamefont {Hagen}},
  \bibinfo {author} {\bibfnamefont {C.~B.}\ \bibnamefont {Lang}}, \bibinfo
  {author} {\bibfnamefont {M.}~\bibnamefont {Limmer}}, \bibinfo {author}
  {\bibfnamefont {D.}~\bibnamefont {Mohler}}, \ and\ \bibinfo {author}
  {\bibfnamefont {A.}~\bibnamefont {Sch{{\"a}}fer}},\ }\href {\doibase
  10.1103/PhysRevD.79.054501} {\bibfield  {journal} {\bibinfo  {journal} {Phys.
  Rev. D}\ }\textbf {\bibinfo {volume} {79}},\ \bibinfo {pages} {054501}
  (\bibinfo {year} {2009})},\ \Eprint {http://arxiv.org/abs/0812.1681}
  {arXiv:0812.1681 [hep-lat]} \BibitemShut {NoStop}%
\bibitem [{\citenamefont {Engel}\ \emph {et~al.}(2010)\citenamefont {Engel},
  \citenamefont {Lang}, \citenamefont {Limmer}, \citenamefont {Mohler},\ and\
  \citenamefont {Sch{{\"a}}fer}}]{Engel:2010my}%
  \BibitemOpen
  \bibfield  {author} {\bibinfo {author} {\bibfnamefont {G.~P.}\ \bibnamefont
  {Engel}}, \bibinfo {author} {\bibfnamefont {C.~B.}\ \bibnamefont {Lang}},
  \bibinfo {author} {\bibfnamefont {M.}~\bibnamefont {Limmer}}, \bibinfo
  {author} {\bibfnamefont {D.}~\bibnamefont {Mohler}}, \ and\ \bibinfo {author}
  {\bibfnamefont {A.}~\bibnamefont {Sch{{\"a}}fer}},\ }\href@noop {} {\bibfield  {journal}
  {\bibinfo  {journal} {Phys. Rev. D}\ }\textbf {\bibinfo {volume} {82}},\
  \bibinfo {pages} {034505} (\bibinfo {year} {2010})},\ \Eprint
  {http://arxiv.org/abs/1005.1748} {arXiv:1005.1748 [hep-lat]} \BibitemShut
  {NoStop}%
\bibitem [{\citenamefont {G{\"u}sken}\ \emph {et~al.}(1989)\citenamefont
  {G{\"u}sken} \emph {et~al.}}]{Gusken:1989ad}%
  \BibitemOpen
  \bibfield  {author} {\bibinfo {author} {\bibfnamefont {S.}~\bibnamefont
  {G{\"u}sken}} \emph {et~al.},\ }\href {\doibase
  10.1016/S0370-2693(89)80034-6} {\bibfield  {journal} {\bibinfo  {journal}
  {Phys. Lett. B}\ }\textbf {\bibinfo {volume} {227}},\ \bibinfo {pages} {266}
  (\bibinfo {year} {1989})}\BibitemShut {NoStop}%
\bibitem [{\citenamefont {Best}\ \emph {et~al.}(1997)\citenamefont {Best},
  \citenamefont {G{\"o}ckeler}, \citenamefont {Horsley}, \citenamefont
  {Ilgenfritz}, \citenamefont {Perlt}, \citenamefont {Rakow}, \citenamefont
  {Sch{{\"a}}fer}, \citenamefont {Schierholz}, \citenamefont {Schiller},\ and\
  \citenamefont {Schramm}}]{Best:1997qp}%
  \BibitemOpen
  \bibfield  {author} {\bibinfo {author} {\bibfnamefont {C.}~\bibnamefont
  {Best}}, \bibinfo {author} {\bibfnamefont {M.}~\bibnamefont {G{\"o}ckeler}},
  \bibinfo {author} {\bibfnamefont {R.}~\bibnamefont {Horsley}}, \bibinfo
  {author} {\bibfnamefont {E.-M.}\ \bibnamefont {Ilgenfritz}}, \bibinfo
  {author} {\bibfnamefont {H.}~\bibnamefont {Perlt}}, \bibinfo {author}
  {\bibfnamefont {P.}~\bibnamefont {Rakow}}, \bibinfo {author} {\bibfnamefont
  {A.}~\bibnamefont {Sch{{\"a}}fer}}, \bibinfo {author} {\bibfnamefont
  {G.}~\bibnamefont {Schierholz}}, \bibinfo {author} {\bibfnamefont
  {A.}~\bibnamefont {Schiller}}, \ and\ \bibinfo {author} {\bibfnamefont
  {S.}~\bibnamefont {Schramm}},\ }\href {\doibase 10.1103/PhysRevD.56.2743}
  {\bibfield  {journal} {\bibinfo  {journal} {Phys. Rev. D}\ }\textbf {\bibinfo
  {volume} {56}},\ \bibinfo {pages} {2743} (\bibinfo {year} {1997})},\ \Eprint
  {http://arxiv.org/abs/hep-lat/9703014} {arXiv:hep-lat/9703014} \BibitemShut
  {NoStop}%
\bibitem [{\citenamefont {L{\"u}scher}\ and\ \citenamefont
  {Wolff}(1990)}]{Luscher:1990ck}%
  \BibitemOpen
  \bibfield  {author} {\bibinfo {author} {\bibfnamefont {M.}~\bibnamefont
  {L{\"u}scher}}\ and\ \bibinfo {author} {\bibfnamefont {U.}~\bibnamefont
  {Wolff}},\ }\href@noop {} {\bibfield  {journal} {\bibinfo  {journal} {Nucl.
  Phys. B}\ }\textbf {\bibinfo {volume} {339}},\ \bibinfo {pages} {222}
  (\bibinfo {year} {1990})}\BibitemShut {NoStop}%
\bibitem [{\citenamefont {Michael}(1985)}]{Michael:1985ne}%
  \BibitemOpen
  \bibfield  {author} {\bibinfo {author} {\bibfnamefont {C.}~\bibnamefont
  {Michael}},\ }\href {\doibase 10.1016/0550-3213(85)90297-4} {\bibfield
  {journal} {\bibinfo  {journal} {Nucl. Phys. B}\ }\textbf {\bibinfo {volume}
  {259}},\ \bibinfo {pages} {58} (\bibinfo {year} {1985})}\BibitemShut
  {NoStop}%
\bibitem [{\citenamefont {Burch}\ \emph {et~al.}(2004)\citenamefont {Burch},
  \citenamefont {Gattringer}, \citenamefont {Glozman}, \citenamefont {Kleindl},
  \citenamefont {Lang},\ and\ \citenamefont {Sch{{\"a}}fer}}]{Burch:2004he}%
  \BibitemOpen
  \bibfield  {author} {\bibinfo {author} {\bibfnamefont {T.}~\bibnamefont
  {Burch}}, \bibinfo {author} {\bibfnamefont {C.}~\bibnamefont {Gattringer}},
  \bibinfo {author} {\bibfnamefont {L.~Y.}\ \bibnamefont {Glozman}}, \bibinfo
  {author} {\bibfnamefont {R.}~\bibnamefont {Kleindl}}, \bibinfo {author}
  {\bibfnamefont {C.~B.}\ \bibnamefont {Lang}}, \ and\ \bibinfo {author}
  {\bibfnamefont {A.}~\bibnamefont {Sch{{\"a}}fer}},\ }\href@noop {} {\bibfield
   {journal} {\bibinfo  {journal} {Phys. Rev. D}\ }\textbf {\bibinfo {volume}
  {70}},\ \bibinfo {pages} {054502} (\bibinfo {year} {2004})},\ \Eprint
  {http://arxiv.org/abs/hep-lat/0405006} {arXiv:hep-lat/0405006} \BibitemShut
  {NoStop}%
\bibitem [{\citenamefont {Lehoucq}\ \emph {et~al.}(1998)\citenamefont
  {Lehoucq}, \citenamefont {Sorensen},\ and\ \citenamefont
  {Yang}}]{Lehoucq:1998xx}%
  \BibitemOpen
  \bibfield  {author} {\bibinfo {author} {\bibfnamefont {R.~B.}\ \bibnamefont
  {Lehoucq}}, \bibinfo {author} {\bibfnamefont {D.~C.}\ \bibnamefont
  {Sorensen}}, \ and\ \bibinfo {author} {\bibfnamefont {C.}~\bibnamefont
  {Yang}},\ }\href@noop {} {\emph {\bibinfo {title} {{ARPACK} {U}sers' {G}uide:
  Solution of large-scale eigenvalue problems with implicitly restarted Arnoldi
  methods}}}\ (\bibinfo  {publisher} {SIAM},\ \bibinfo {address} {New York},\
  \bibinfo {year} {1998})\BibitemShut {NoStop}%
\bibitem [{\citenamefont {Gattringer}\ \emph {et~al.}(2008)\citenamefont
  {Gattringer}, \citenamefont {Glozman}, \citenamefont {Lang}, \citenamefont
  {Mohler},\ and\ \citenamefont {Prelovsek}}]{Gattringer:2008be}%
  \BibitemOpen
  \bibfield  {author} {\bibinfo {author} {\bibfnamefont {C.}~\bibnamefont
  {Gattringer}}, \bibinfo {author} {\bibfnamefont {L.~Y.}\ \bibnamefont
  {Glozman}}, \bibinfo {author} {\bibfnamefont {C.~B.}\ \bibnamefont {Lang}},
  \bibinfo {author} {\bibfnamefont {D.}~\bibnamefont {Mohler}}, \ and\ \bibinfo
  {author} {\bibfnamefont {S.}~\bibnamefont {Prelovsek}},\ }\href {\doibase
  10.1103/PhysRevD.78.034501} {\bibfield  {journal} {\bibinfo  {journal} {Phys.
  Rev. D}\ }\textbf {\bibinfo {volume} {78}},\ \bibinfo {pages} {034501}
  (\bibinfo {year} {2008})},\ \Eprint {http://arxiv.org/abs/0802.2020}
  {arXiv:0802.2020 [hep-lat]} \BibitemShut {NoStop}%
\bibitem [{\citenamefont {Engel}\ \emph {et~al.}(2012)\citenamefont {Engel},
  \citenamefont {Lang}, \citenamefont {Limmer}, \citenamefont {Mohler},\ and\
  \citenamefont {Sch{\"a}fer}}]{Engel:2011aa}%
  \BibitemOpen
  \bibfield  {author} {\bibinfo {author} {\bibfnamefont {G.~P.}\ \bibnamefont
  {Engel}}, \bibinfo {author} {\bibfnamefont {C.~B.}\ \bibnamefont {Lang}},
  \bibinfo {author} {\bibfnamefont {M.}~\bibnamefont {Limmer}}, \bibinfo
  {author} {\bibfnamefont {D.}~\bibnamefont {Mohler}}, \ and\ \bibinfo {author}
  {\bibfnamefont {A.}~\bibnamefont {Sch{\"a}fer}},\ }\href@noop {} {\bibfield
  {journal} {\bibinfo  {journal} {Phys. Rev. D}\ }\textbf {\bibinfo {volume}
  {85}},\ \bibinfo {pages} {034508} (\bibinfo {year} {2012})},\ \Eprint
  {http://arxiv.org/abs/1112.1601} {arXiv:1112.1601 [hep-lat]} \BibitemShut
  {NoStop}%
\bibitem [{\citenamefont {De~Rujula}\ \emph {et~al.}(1975)\citenamefont
  {De~Rujula}, \citenamefont {Georgi},\ and\ \citenamefont
  {Glashow}}]{De_Rujula:1975ge}%
  \BibitemOpen
  \bibfield  {author} {\bibinfo {author} {\bibfnamefont {A.}~\bibnamefont
  {De~Rujula}}, \bibinfo {author} {\bibfnamefont {H.}~\bibnamefont {Georgi}}, \
  and\ \bibinfo {author} {\bibfnamefont {S.~L.}\ \bibnamefont {Glashow}},\
  }\href {\doibase 10.1103/PhysRevD.12.147} {\bibfield  {journal} {\bibinfo
  {journal} {Phys. Rev. D}\ }\textbf {\bibinfo {volume} {12}},\ \bibinfo
  {pages} {147} (\bibinfo {year} {1975})}\BibitemShut {NoStop}%
\bibitem [{\citenamefont {Isgur}\ and\ \citenamefont
  {Karl}(1978)}]{Isgur:1978xj}%
  \BibitemOpen
  \bibfield  {author} {\bibinfo {author} {\bibfnamefont {N.}~\bibnamefont
  {Isgur}}\ and\ \bibinfo {author} {\bibfnamefont {G.}~\bibnamefont {Karl}},\
  }\href {\doibase 10.1103/PhysRevD.18.4187} {\bibfield  {journal} {\bibinfo
  {journal} {Phys. Rev. D}\ }\textbf {\bibinfo {volume} {18}},\ \bibinfo
  {pages} {4187} (\bibinfo {year} {1978})}\BibitemShut {NoStop}%
\bibitem [{\citenamefont {Glozman}\ and\ \citenamefont
  {Riska}(1996)}]{Glozman:1995fu}%
  \BibitemOpen
  \bibfield  {author} {\bibinfo {author} {\bibfnamefont {L.~Y.}\ \bibnamefont
  {Glozman}}\ and\ \bibinfo {author} {\bibfnamefont {D.~O.}\ \bibnamefont
  {Riska}},\ }\href {\doibase 10.1016/0370-1573(95)00062-3} {\bibfield
  {journal} {\bibinfo  {journal} {Phys. Rept.}\ }\textbf {\bibinfo {volume}
  {268}},\ \bibinfo {pages} {263} (\bibinfo {year} {1996})},\ \Eprint
  {http://arxiv.org/abs/hep-ph/9505422} {arXiv:hep-ph/9505422} \BibitemShut
  {NoStop}%
\end{thebibliography}
\end{document}